\documentclass[preprint,12pt,review]{elsarticle}

\usepackage{amssymb}
\usepackage{amsmath}
\usepackage{subfigure}
\usepackage{wrapfig}
\usepackage{booktabs}
\usepackage[table]{xcolor}
\definecolor{skyblue}{HTML}{87CEEB}
\usepackage{amsmath,amsfonts}
\usepackage{algorithmic}
\usepackage{algorithm}
\usepackage{array}
\usepackage{color}
\usepackage{multirow}
\usepackage{url}
\usepackage{hyperref}

\begin{document}

\begin{frontmatter}



\title{Balancing User Preferences by Social Networks: A Condition-Guided Social Recommendation Model for Mitigating Popularity Bias}

\author[SAI]{Xin He}
\author[PLU]{Wenqi Fan}
\author[SAI]{Ruobing Wang}
\author[SAI]{Yili Wang}
\author[CS]{Ying Wang}
\author[GU]{Shirui Pan}
\author[SAI]{Xin Wang}
\affiliation[SAI]{organization={ The Department
of Computer Science and Technology, the School of Artificial Intelligence (SAI), at Jilin University},
            addressline={2699 Qianjin Street},
            city={Chang Chun},
            postcode={130000},
            state={Jilin},
            country={China}}

\affiliation[PLU]{organization={The Hong Kong Polytechnic University},
            addressline={Kowloon},
            city={Hong Kong},
            postcode={000000},
            state={Hong Kong},
            country={China}}

\affiliation[CS]{organization={The Department of Computer Science and Technology,
College of Computer Science and Technology, at Jilin University},
            addressline={2699 Qianjin Street},
            city={Chang Chun},
            postcode={130000},
            state={Jilin},
            country={China}}

\affiliation[GU]{organization={Griffith University},
            addressline={Parklands Drive},
            city={Southport},
            postcode={4222},
            state={Queensland},
            country={Australia}}



\begin{abstract}
Social recommendation models weave social interactions into their design to provide uniquely personalized recommendation results for users.
However, social networks not only amplify the popularity bias in recommendation models, resulting in more frequent recommendation of hot items and fewer long-tail items, but also include a substantial amount of redundant information that is essentially meaningless for the model's performance.
Existing social recommendation models often integrate the entire social network directly, with little effort to filter or adjust social information to mitigate popularity bias introduced by the social network.
In this paper, we propose a Condition-Guided Social Recommendation Model (named CGSoRec) to mitigate the model's popularity bias by denoising the social network and adjusting the weights of user's social preferences.
More specifically, CGSoRec first includes a Condition-Guided Social Denoising Model (CSD) to remove redundant social relations in the social network for capturing users' social preferences with items more precisely. 
Then, CGSoRec calculates users' social preferences based on denoised social network and adjusts the weights in users' social preferences to make them can counteract the popularity bias present in the recommendation model.
At last, CGSoRec includes a Condition-Guided Diffusion Recommendation Model (CGD) to introduce the adjusted social preferences as conditions to control the recommendation results for a debiased direction.
Comprehensive experiments on three real-world datasets demonstrate the effectiveness of our proposed method.
The code is in: \href{https://github.com/hexin5515/CGSoRec}{https://github.com/hexin5515/CGSoRec}. 
\end{abstract}




\end{frontmatter}


\section{Introduction}
Over the past decade, social recommendation has become a vital component of recommendation systems. 
Recent studies~\cite{yu2021self,quan2023robust,lacker2024socially,ma2024madm,wu2022disentangled} have demonstrated that incorporating user-user social relationships can enhance models' ability to effectively capture collaborative signals between users and items in certain scenarios. 
This improvement is based on the theory of social homogeneity~\cite{mcpherson2001birds,zheng2023rumor}, which posits that users with a social relationship are likely to share similar preferences.
Social homogeneity has been corroborated in a diverse range of application domains within recommendation systems~\cite{sha2022wants,dong2020automated,xi2021deep}.


\begin{figure*}[h]
\centering

{\subfigure
{\includegraphics[width=0.24\linewidth]{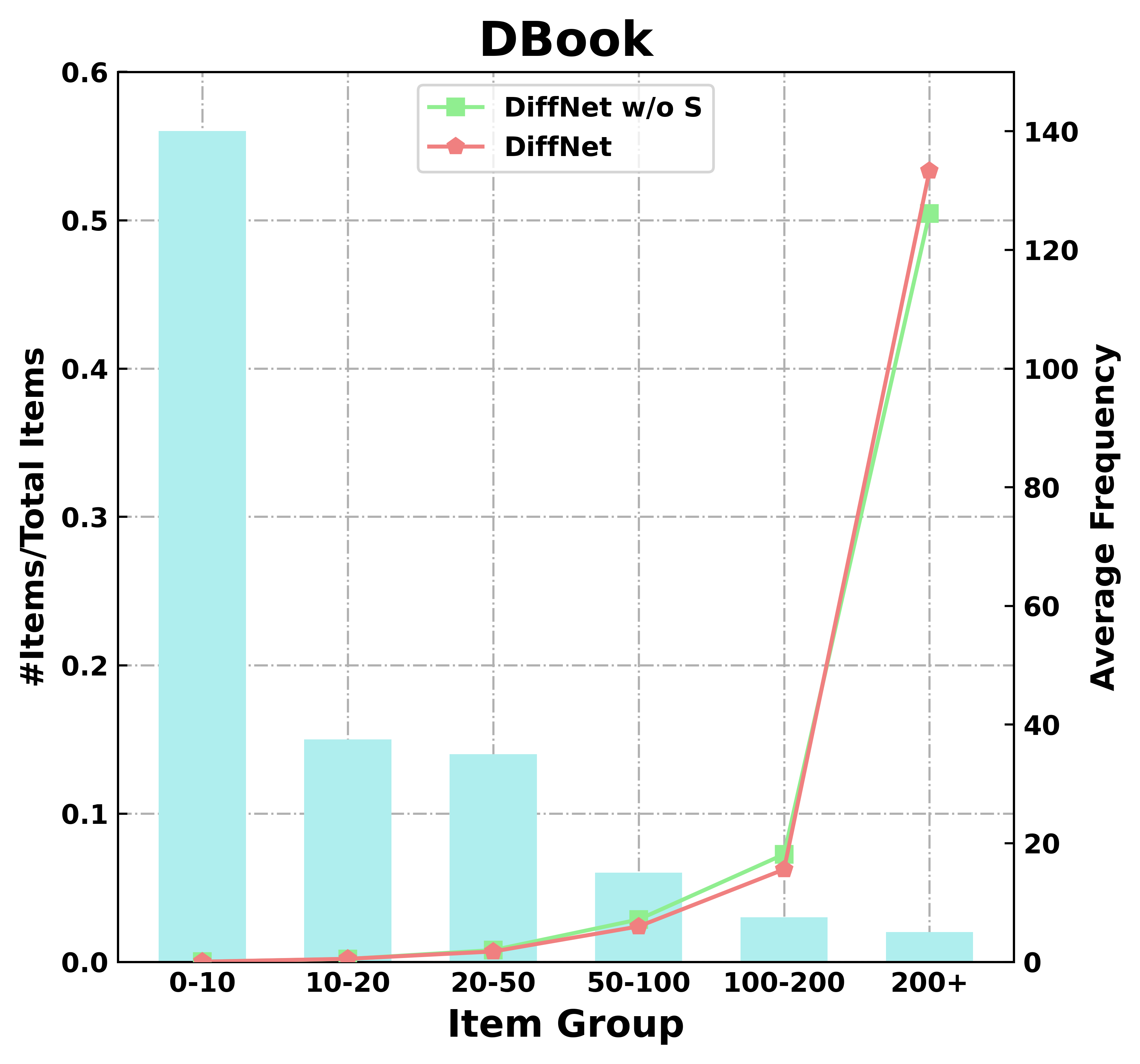}}}
{\subfigure
{\includegraphics[width=0.24\linewidth]{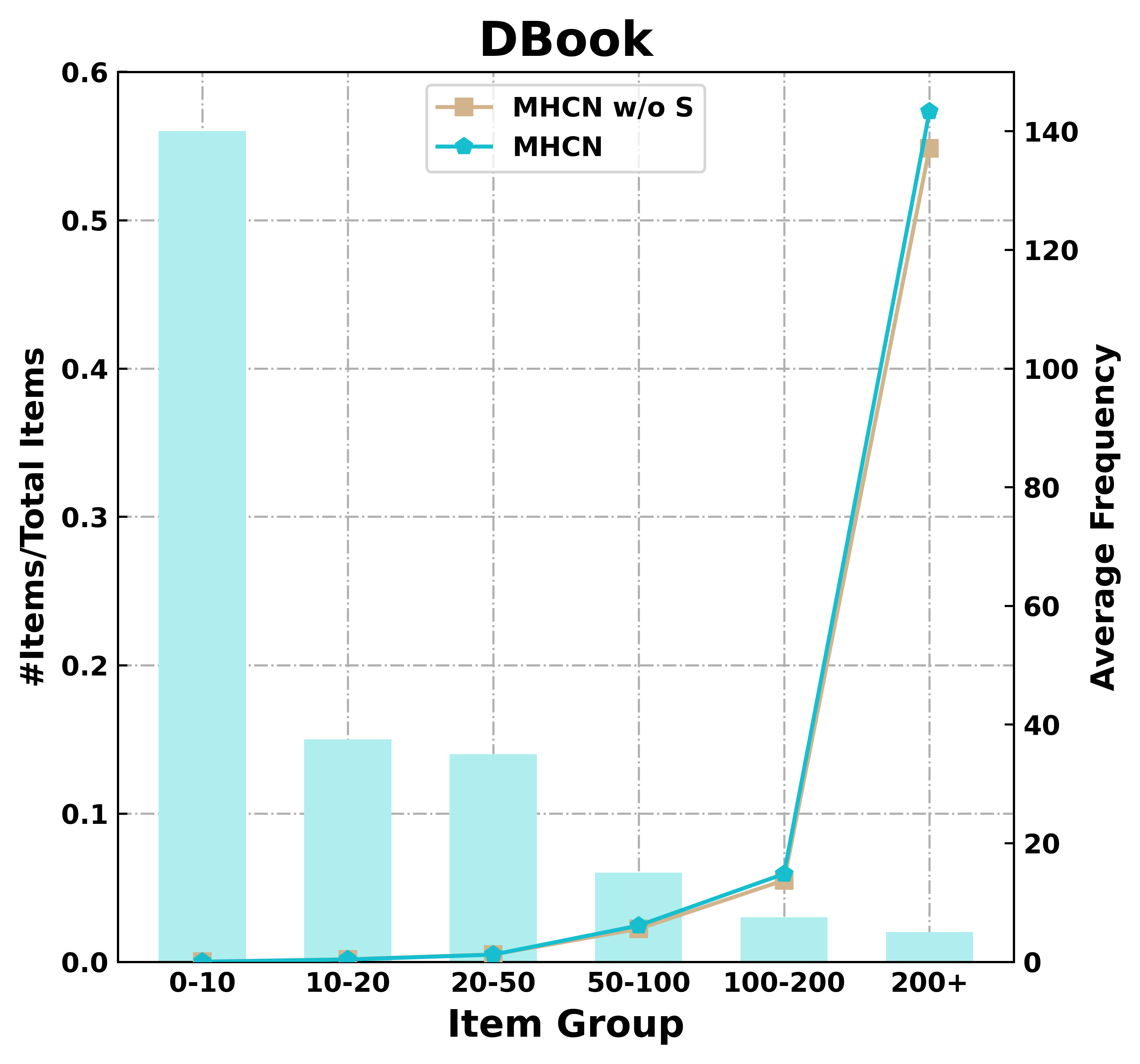}}}
{\subfigure
{\includegraphics[width=0.24\linewidth]{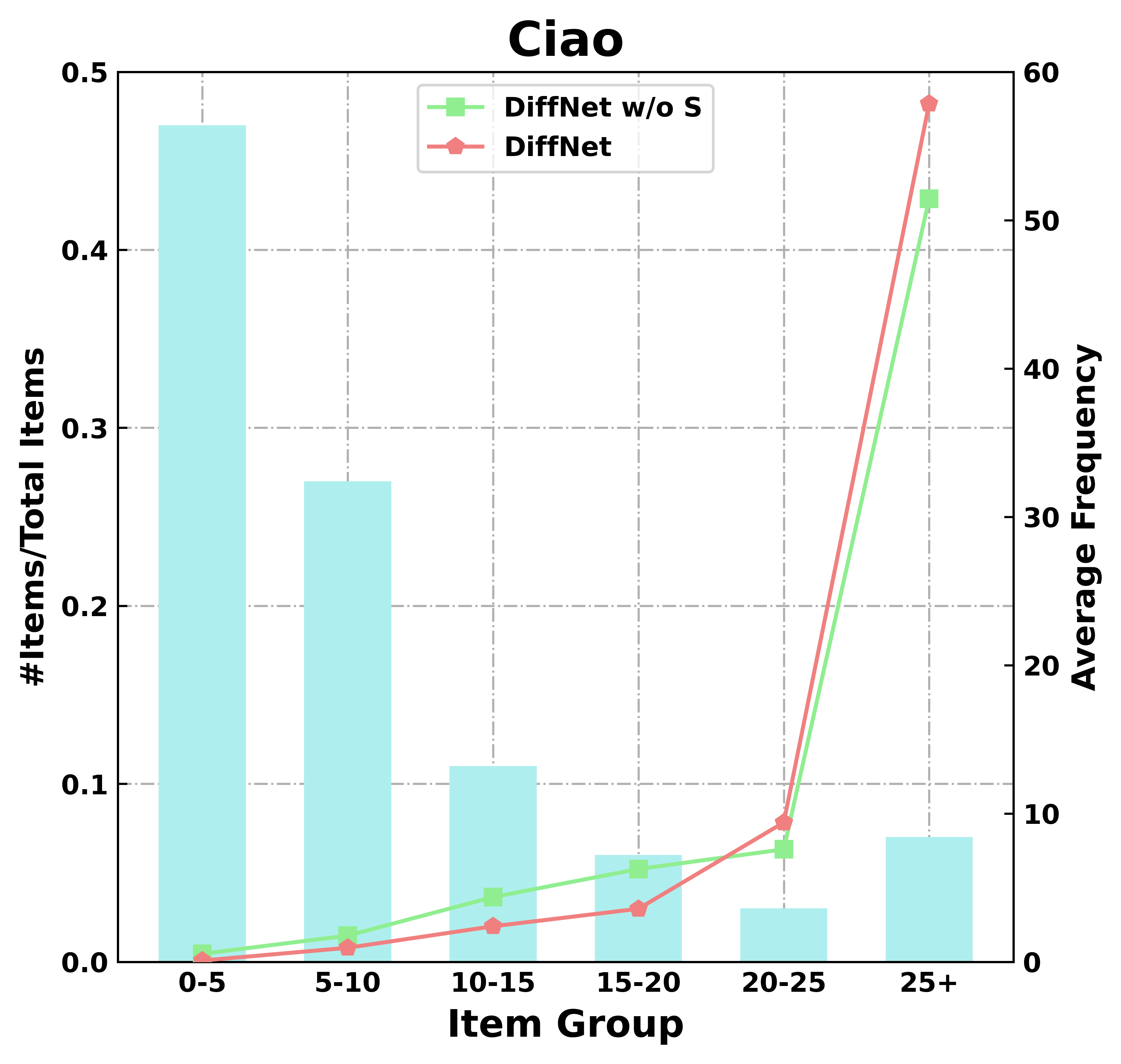}}}
{\subfigure
{\includegraphics[width=0.24\linewidth]{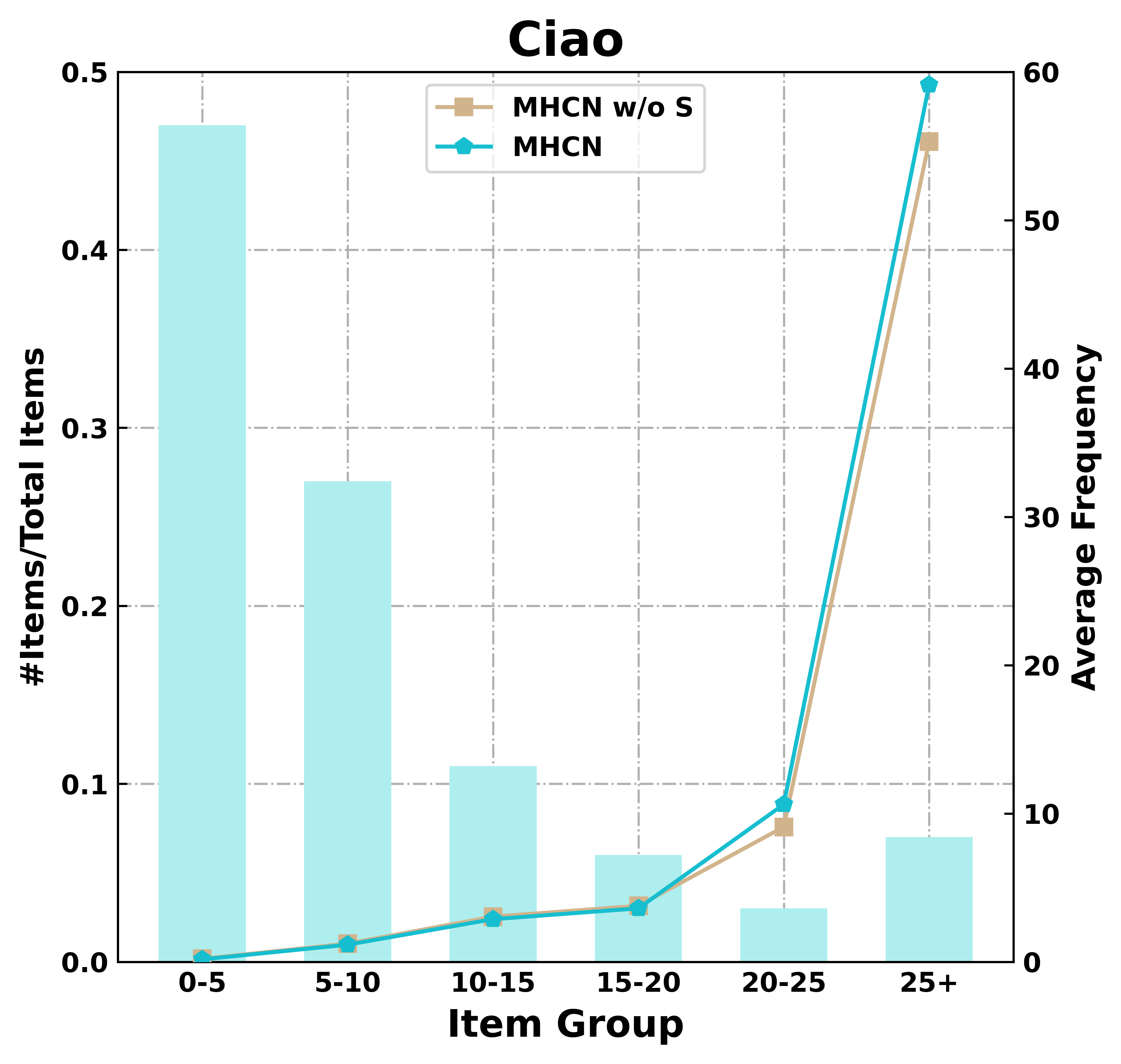}}}%

\caption{Popularity bias comparison between introduced and removed Social Network versions of DiffNet and MHCN model.}\label{popular_bias_between_DiffNet_and_DiffNet_w/o_S}
\end{figure*}




Despite their superior effectiveness, a key limitation of social recommendation models is that they tend to amplify popularity bias~\cite{wei2021model} which makes popular items receive more recommendation. 
Fig.~\ref{popular_bias_between_DiffNet_and_DiffNet_w/o_S} presents that the social recommendation methods (Diffnet~\cite{wu2019neural} and MHCN~\cite{yu2021self}) lead to amplifying popularity bias on a real-world Ciao and DBook dataset~\cite{tang2012mtrust}. 
We train two classical social recommendation models and quantify item frequencies in the top-$K$ recommendation across all users~\cite{wei2021model}.
As can be seen, compared to the versions DiffNet (MHCN) w/o S that removes social information, DiffNet (MHCN) increases the frequency of recommendation for hot items.
This exacerbation may enhance the performance of social recommendation models on traditional recommendation datasets, but in scenarios requiring fair recommendation~\cite{wei2021model,zheng2021disentangling}, it often leads to a decrease in model performance.
Nevertheless, there is still no research focusing on specifically designing models to mitigate popularity bias in social recommendation.
This motivates us to pose the following research question: \textbf{how to design a method to mitigate the popularity bias due to the introduction of social relationships?}

\begin{wrapfigure}{r}{6.5cm} 
    \includegraphics[width=6cm]{ 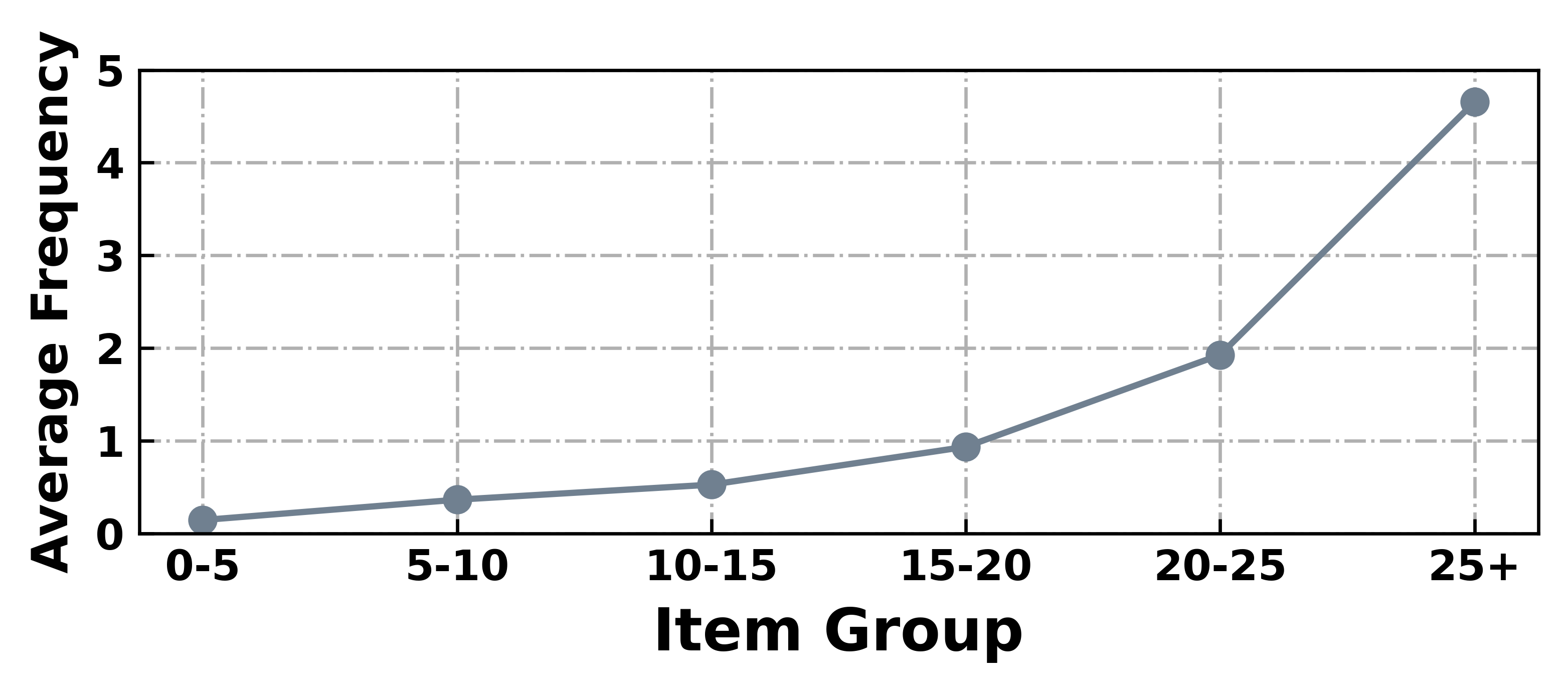}\
    \caption{Average frequency of items recommended based on a user's social preferences.}
\label{Average_frequency_of_a_user's_social_preferences}
\end{wrapfigure}

To answer this question, we first examine why social networks amplify the popularity bias in social recommendation models.
We quantify the distribution of items favored by a user's social neighbors, which named the user's social preference~\cite{li2019social}.
As shown in Fig.~\ref{Average_frequency_of_a_user's_social_preferences}, in a users' social preferences, hot items are recommended more often than long-tail items, exhibiting a trend similar to the model's inherent popularity bias shown in Fig.~\ref{popular_bias_between_DiffNet_and_DiffNet_w/o_S}.
Therefore, the integration of social networks into the recommendation model further amplifies its popularity bias, leading to an increased recommendation of hot items by the model.
It is important to note that users' social preferences for long-tail items are also present but less pronounced compared to their preferences for hot items.
Therefore, inspired by knowledge editing method~\cite{orgad2023editing,meng2022locating,de2021editing} in the field of Computer Vision (CV) and Natural Language Processing (NLP), we propose to adjust the weights in user social preferences. This adjustment aims to reveal preferences that counteract the inherent bias within the recommendation model, thereby mitigating its popularity bias.

Recently, with the development of diffusion models in the field of recommendation systems, some studies~\cite{wang2023diffusion,yang2024generate,walker2022recommendation,choi2023blurring} have shown that diffusion recommendation models can model the complex interactions between users and items more accurately compared to traditional matrix factorization recommendation models and graph neural network recommendation models.
Besides, with the development of Condition-Guided Diffusion Models in the fields of CV and NLP, numerous studies~\cite{tashiro2021csdi,dhariwal2021diffusion,zhang2023adding,ho2022classifier,yu2023freedom} have demonstrated that Condition-Guided Diffusion Models can proficiently modulate generation results as desired.
This motivates our investigation into the generation of unbiased recommendation results through the utilization of a condition-guided diffusion recommendation model.
However, this investigation also presents us with two distinct challenges: 
(1) The incorporation of social information often exacerbates the popularity bias inherent in recommendation models. 
Therefore, determining how to adjust users' social preferences when incorporating social information as a condition into the recommendation model is crucial for mitigating the popularity bias.
(2) Nearly all social networks contain an abundance of redundant relations~\cite{quan2023robust,yu2020enhancing,wang2021denoising}, which are not only irrelevant for improving the performance of the social recommendation models but also detrimentally affect their performance. 
Therefore, it is critical to remove these redundant social relations in social networks when taking advantage of social information.

To address the above issues, in this paper, we propose a Condition-Guided Social Recommendation Model (named \textbf{CGSoRec}), which can mitigate the popularity bias by conditionally steering the recommendation results in alignment with the user's social preferences. 
More specifically, we first propose a Condition-Guided Social Denoising Model (CSD) to remove redundant or noisy social relations in social networks to reflect users' social preferences more accurately.
Then, we derive users' social preferences based on denoised social networks and adjust the weights within these preferences to counteract the inherent popularity bias in the model.
At last, we propose a Condition-Guided Diffusion Recommendation Model (CGD) to introduce users' social preferences into diffusion recommendation models as conditions and mitigate the popularity bias by adjusting the weights of hot and long-tail items within user's social preferences. 

In summary, this work makes the following contributions:
\begin{itemize}
    \item We are the first to address the issue of popularity bias in social recommendation models by incorporating adjusted social preferences as conditions in a Condition-Guided Diffusion Recommendation Model (CGD). 
    \item We innovatively propose a Condition-Guided Social Denoising Model (CSD) to eliminate redundant information in social networks, thereby enabling the model to capture users' social preferences more accurately.
    \item We conduct comprehensive experiments and ablation studies on three real-world datasets to demonstrate the effectiveness of our proposed method.
\end{itemize}

\section{Related Work}
\subsection{Social Recommendation Model}
With the development of social media, leveraging social relations for recommendation has garnered considerable interest in the field in recent years~\cite{jamali2010matrix,wu2019neural,fan2019graph,guo2020ipgan,fan2019deep}.
More specifically, existing social recommendation models leverages various techniques~\cite{he2020lightgcn,wu2021self_survey,sheu2021knowledge} to integrate social networks into their framework, aiming to enhance the overall performance of these models.
For instance, GraphRec\cite{fan2019graph} utilizes the capabilities of graph neural networks to improve the representation learning of users and items within the context of social recommendation. 
DiffNet~\cite{wu2019neural} employs a GNNs model to simulate the process of social influence propagation, thereby achieving more refined representations of users and items. 
MHCN~\cite{yu2020enhance} integrates self-supervised learning and hypergraph convolution techniques to augment relational data, focusing on encoding higher-order and complex connection patterns, thereby significantly advancing the efficacy of social recommendation models.
\subsection{Generative Recommendation Model}
Over the past decade, there has been a rising fascination with generative models such as Variational Auto-Encoders (VAEs)~\cite{kingma2013auto,li2024disentangled,li2023distvae} and Denoising Diffusion Probabilistic Model (DDPM)~\cite{ho2020denoising,xuan2024diffusion,wang2024conditional} for personalized recommendation due to their remarkable capacity of modeling non-linear user-item interactions and denoising.
For example, Mult-VAE~\cite{liang2018variational} proposes a variant of vae for collaborative filtering on implicit feedback data, which enable to go beyond linear factor models with limited modeling capacity. 
DiffRec~\cite{wang2023diffusion} proposes a new recommendation paradigm for generative recommendation models and improves the performance of the model through a time-aware reweighting strategy. 
\subsection{Condition-Guided Diffusion Model}
Meanwhile, with the development of conditional diffusion model, numerous studies~\cite{ho2022classifier,dhariwal2021diffusion,zhou2023sparsefusion,zhong2023guided} have demonstrated the advantages of employing conditional diffusion models to improve the quality of the generated results. 
For example, the classifier-guidance diffusion model~\cite{dhariwal2021diffusion} employs a conditional classifier to direct the diffusion process, facilitating the creation of outputs that are more precise and aligned with specified conditions. 
The classifier-free diffusion model~\cite{ho2022classifier} incorporates conditional guidance to steer the generation process, maintaining its training procedure unchanged while allowing for the creation of diverse results guided by specific conditions. 
In this work, we explore how to denoise social networks and utilize the adjusted user social preference to control the recommendation results of the diffusion recommendation model for mitigating the popularity bias.

\section{Preliminaries}
In this section, we introduce definitions and notations that are used throughout the paper.
Let $U = \{u_1, u_2, ..., u_m\}$ and $V = \{v_1, v_2, ..., v_n\}$ be the sets of users and items, respectively, where $m$ is the number of users and $n$ is the number of items.
We denote $\mathbf{R}\in\mathbb{R}^{m\times{n}}$ as a user-item interaction matrix, where $r_{ij}$ = 1 if user $u_i$ has interacted with item $v_j$, otherwise $r_{ij}$ = 0. 
We denote $\mathbf{S}\in\mathbb{R}^{m\times{m}}$ as a user-user social matrix, where $s_{ii'} = 1$ if user $u_i$ has a social relation with user $u_{i'}$, otherwise $s_{ii'} = 0$.
In this work, since the data in the social network cannot be injected directly into the user's interaction vector $x_{i}$, we utilize the user social preference matrix $\mathbf{R}^{s} \in\mathbb{R}^{m\times{n}} = \mathbf{S}\mathbf{R}$ to enhance the performance of the recommendation model.
The row vector ${\mathbf{x}_i}^{s}=[x_{i1}^{s}, x_{i2}^{s}, ..., x_{in}^{s}]$ in the matrix $\mathbf{R}^{s}$ denotes the interaction of the target user's social neighborhood with all items. 
Given $\mathbf{R}$ and $\mathbf{R}^{s}$, our goal is to direct the model's denoising of interaction matrix $\mathbf{R}$ using the social preference matrix $\mathbf{R}^{s}$ for predicting users’ unknown preferences towards items and mitigate the popularity bias.

\section{The Proposed Method}

In this section, we first introduce our proposed Condition-Guided Social Recommendation Model (\textbf{CGSoRec}). Then, we provide a comprehensive overview of the model training process and inferencing process.

\subsection{An Overview of the Proposed Method}
\begin{figure*}[t]
  \centering
  \includegraphics[width=1\textwidth]{ 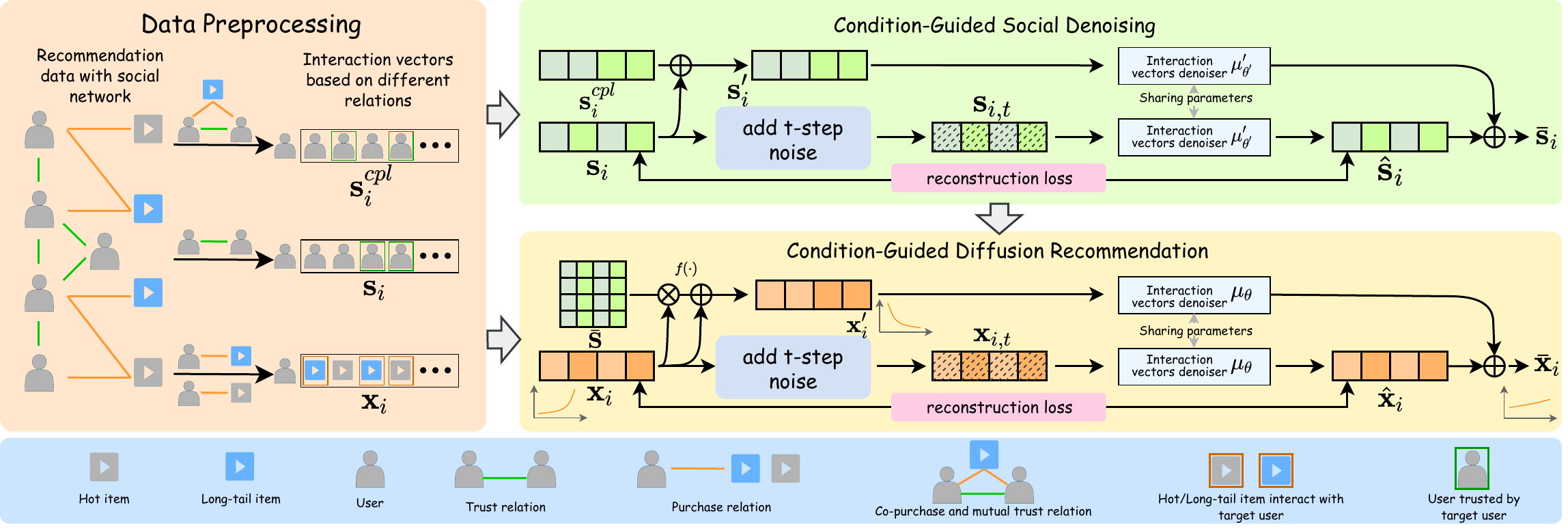}
  \caption{The overall framework of the proposed method. 
  } 
\label{overall framework}
\end{figure*}

The overall architecture of the proposed model is shown in  Fig.~\ref{overall framework}, which consists of two main submodels:
(1) \textbf{Condition-Guided Social Denoising Model (CSD)}, which is introduced to reflect users’ social preference more accurately by introducing user-user interaction for co-purchasing long-tail items as conditions and removing redundant or noisy social relationships in social networks;
(2) \textbf{Condition-Guided Diffusion Recommendation Model (CGD)}, which aims to mitigate the popularity bias of the recommendation model by incorporating the social information as a condition into the diffusion recommendation model and adjusting the weights of hot/long-tail items in the user's social preference.
Specifically, we first derive the user-item interaction vector $\mathbf{x}_{i}$, as well as the user-user interaction matrices $\mathbf{s}^{cpl}_{i}$ and $\mathbf{s}_{i}$, based on varying user-item and user-user relationships. Then, we compute the interaction vector $\mathbf{s}_{i}'$ to direct the CSD in the denoising process of the social interaction vector $\mathbf{s}_{i}$. At last, utilizing the denoised user social interaction vector $\bar{\mathbf{s}}_{i}$, we calculate the user's social preferences and apply these preferences, adjusted for weight as $\mathbf{x}_{i}'$, to steer the CGD, thereby achieving bias-reduced recommendation outcomes $\bar{\mathbf{x}}_{i}$.
Next, we will first introduce the general paradigm of the diffusion recommendation model, followed by a detailed description of the two submodels.

\subsection{Diffusion Recommendation Model}
As one of the most representative diffusion recommendation methods, DiffRec~\cite{wang2023diffusion} effectively fills in the missing values in user-item interaction vectors in a denoising manner, thereby providing personalized recommendations for users. Therefore,  we adopt DiffRec as our target encoder for denoising user-item interaction vectors in user-item interactions matrix $\mathbf{R}\in\mathbb{R}^{m\times{n}}$.
The forward diffusion process of the diffusion recommendation model can be formulated as follows:
\begin{align}
        q(\mathbf{x}_{i,t}|\mathbf{x}_{i,t-1}) = \mathcal{N}(\mathbf{x}_{i,t};\sqrt{1-\beta_{t}}\mathbf{x}_{i,t-1}, \beta_{t} \mathbf{I}),
\end{align}where $\beta_{t}$ controls the strength of the Gaussian noise added at each step of the forward diffusion process. We can directly obtain
$\mathbf{x}_{i,t}$ from $\mathbf{x}_{i,0}$ due to the additivity
of two independent Gaussian noises~\cite{luo2022understanding} as:
\begin{align}\label{fdp}
    q(\mathbf{x}_{i,t}|\mathbf{x}_{i,0}) = \mathcal{N}(\mathbf{x}_{i,t};\sqrt{\overline{\alpha}_{i,t}}\mathbf{x}_{i,0}, (1-\overline{\alpha}_{t}) \mathbf{I}),
\end{align}where $\alpha_{t}=1-\beta_{t}$, $\overline{\alpha}_{t}=\prod_{t'=1}^{t}\alpha_{t'}$ and we can directly obtain $\mathbf{x}_{i,t}$ as: $\mathbf{x}_{i,t}=\sqrt{\overline{\alpha}_{t}}\mathbf{x}_{i,0}+\sqrt{1-\overline{\alpha}_{t}}\epsilon$ with $\epsilon \sim \mathcal{N}(0,1)$
. Then, the diffusion recommendation model gradually recovers users’ interactions by the denoising transition step:
\begin{align} \label{diffrec_backward}
    p_{\theta}(\mathbf{x}_{i,t-1}|\mathbf{x}_{i,t}) = \mathcal{N}(\mathbf{x}_{i,t-1};\mu_{\theta}(\mathbf{x}_{i,t},t), \Sigma_{\theta}(\mathbf{x}_{i,t},t)),
\end{align}where $\mu_{\theta}(\mathbf{x}_{i,t},t)$ and $\Sigma_{\theta}(\mathbf{x}_{i,t},t)$ are the Gaussian parameters outputted by any neural networks with learnable parameters $\theta$.
The diffusion recommendation model is mainly optimized by regulating the $p_{\theta}(\mathbf{x}_{i,t-1}|\mathbf{x}_{i,t})$ to approximate tractable ground-truth transition step $q(\mathbf{x}_{i,t-1}|\mathbf{x}_{i,t},\mathbf{x}_{i,0})$ which can be rewritten as~\cite{luo2022understanding}:
\begin{align}
    q(\mathbf{x}_{i,t-1}|\mathbf{x}_{i,t},\mathbf{x}_{i,0})\propto \mathcal{N}(\mathbf{x}_{i,t-1};\Tilde{\mu}(\mathbf{x}_{i,t},\mathbf{x}_{i,0},t),\sigma^{2}(t)\mathbf{I}),
\end{align}
\begin{align}
    \left \{ \begin{array}{l}
         \Tilde{\mu}(\mathbf{x}_{i,t},\mathbf{x}_{i,0},t)=\frac{\sqrt{\alpha_{t}}(1-\overline{\alpha}_{t-1})}{1-\overline{\alpha}_{t}}\mathbf{x}_{i,t} + \frac{\sqrt{\overline{\alpha}_{t-1}}(1-\alpha_{t})}{1-\overline{\alpha}_{t}}\mathbf{x}_{i,0},  \\
         \sigma^{2}(t)= \frac{(1-\alpha_{t})(1-\overline{\alpha}_{t-1})}{1-\overline{\alpha}_{t}},
    \end{array} \right.
\end{align}where $\Tilde{\mu}(\mathbf{x}_{i,t},\mathbf{x}_{i,0},t)$ and $\sigma^{2}(t)$ are the mean and covariance of $q(\mathbf{x}_{i,t-1}|\mathbf{x}_{i,t},\mathbf{x}_{i,0})$ decoupled from Eq.~(\ref{fdp}). Naturally, the optimization objective of the diffusion recommendation model can be changed from forcing $p_{\theta}(\mathbf{x}_{i,t-1}|\mathbf{x}_{i,t})$ to $q(\mathbf{x}_{i,t-1}|\mathbf{x}_{i,t},\mathbf{x}_{i,0})$ into pushing $\mu_{\theta}(\mathbf{x}_{i,t},t)$ be closed to $ \Tilde{\mu}(\mathbf{x}_{i,t},\mathbf{x}_{i,0},t)$. And $\mu_{\theta}(\mathbf{x}_{i,t},t)$ can be formulated as:
\begin{align}
     \mu_{\theta}(\mathbf{x}_{i,t},t)=\frac{\sqrt{\alpha_{t}}(1-\overline{\alpha}_{t-1})}{1-\overline{\alpha}_{t}}\mathbf{x}_{i,t} + \frac{\sqrt{\overline{\alpha}_{t-1}}(1-\alpha_{t})}{1-\overline{\alpha}_{t}}\hat{x}_{\theta}(\mathbf{x}_{i,t},t),
\end{align}where $\hat{x}_{\theta}(\mathbf{x}_{i,t},t)$ is the neural networks with learnable parameters $\theta$.
With the user-item interaction vector $\mathbf{x}_i$, the goal of the diffusion recommendation model is to predict users’ unknown preferences towards items by Eq.~\ref{diffrec_backward}.

\subsection{Condition-Guided Social Denoising Model}
Original social networks often contain excessive redundant information, which may lead to the model being unable to accurately reflect users' social preferences.
Therefore, we propose a Condition-Guided Social Denoising Model (CSD) that removes redundant or noisy social relationships in social networks. 
Similar to the diffusion recommendation model, we consider the row vector $\mathbf{s}_i=[s_{i1}, s_{i2}, ..., s_{im}]$ in the social matrix $\mathbf{S}\in \mathbb{R}^{m\times{m}}$, which encapsulates the user's social interaction records, as the input to CSD, and adds Gaussian noise to $\mathbf{s}_i$ using a similar method as Diffusion Recommendation Model. 
Subsequently, a neural network
$\mu_{\theta'}$ is employed to learn the denoising process, which can be formulated as follows:
\begin{align}
    p_{\theta'}(\mathbf{s}_{i,t-1}|\mathbf{s}_{i,t}) = \mathcal{N}(\mathbf{s}_{i,t-1};\mu_{\theta'}(\mathbf{s}_{i,t},t), \Sigma_{\theta'}(\mathbf{s}_{i,t},t)),
\end{align}where $\theta'$ is the learnable parameters of the Social Denoising Model. 
Then, we introduce co-purchase relationships as a condition in the CSD. 
Specifically, we utilize the user-item interaction matrix $\mathbf{R}_{l}$ which records the user interactions with long-tail items, to derive the user-user interaction matrix $\mathbf{S}^{cpl}$ for co-purchasing long-tail items.
And we also incorporate the user-user interaction matrix $\mathbf{S}^{cpl}$ for co-purchasing long-tail items into the user-user interaction matrix $\mathbf{S}$ as condition so that CSD prioritizes users with similar long-tail item preferences:
\begin{align}\label{obtain_scpl}
    \mathbf{S}^{cpl} = \mathbf{R}_{l} \mathbf{R}_{l}^{T},
\end{align}
\begin{align}\label{obtain_condition_s}
    \mathbf{S}' = \delta \mathbf{J} \odot \mathbf{S}^{cpl} + \mathbf{S},
\end{align}
where $\mathbf{J} \in\mathbb{R}^{m\times{n}}$ is the matrix whose elements are all $1$. $s_{ij}^{cpl}$ denotes the sum of co-purchasing long-tail item relationships from user $u_i$ to user $u_j$. 
The larger value of $s_{ij}^{cpl}$ indicates that user $u_i$ and $u_j$ exhibit identical preferences for long-tail items. 
$\odot$ denotes the element-wise product.
$\delta$ is a coefficient to control the effect of the condition on the model's denoising results. The denoising process can be formulated as follows:
\begin{align}
    p_{\theta'}(\mathbf{s}_{i,t-1}|\mathbf{s}_{i,t}, \mathbf{s}_{i}') = \mathcal{N}(\mathbf{s}_{i,t-1};\mu_{\theta'}'(\mathbf{s}_{i,t},\mathbf{s}_{i}',t), \Sigma_{\theta'}),
\end{align}
\begin{align}\label{s_rdp}
    \mu_{\theta'}'(\mathbf{s}_{i,t},\mathbf{s}_{i}',t)=(1-\eta) \mu_{\theta'}(\mathbf{s}_{i,t},t)+\eta\mu_{\theta'}(\mathbf{s}_{i}',t),
\end{align}where $\mathbf{s}_i'=[s_{i1}', s_{i2}', ..., s_{in}']$ is the row vectors in matrix $\mathbf{S}'$ recording user-user social relationships and user co-purchase relationships for long-tail items.
$\mu_{\theta'}'(\mathbf{s}_{i,t},\mathbf{s}_{i}',t)$ and $\Sigma_{\theta'}=\Sigma_{\theta'}(\mathbf{s}_{i,t},t)$ are the Gaussian parameters by neural networks with learnable parameters $\theta'$.
The hyper-parameter $\eta$ is employed in our CSD to regulate the degree in the inferencing process integrating the user-user co-purchasing long-tail items relationships as condition.
By incorporating the social relationships of users who co-purchase long-tail items as conditions, CSD gains the capability to prioritize the social relationships between users who share preferences for the same long-tail items when removing redundant or noisy social relationships in social networks.

\begin{figure*}[t]
  \centering
  \includegraphics[width=1.0\linewidth]{ 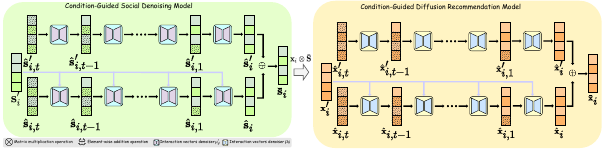}
  \caption{The Joint Inferencing process of the proposed method. 
  } 
\label{joint_inference}
\end{figure*}

\subsection{Condition-Guided Diffusion Recommendation Model}
\subsubsection{Adjusting the weights of items in social preferences}
Instead of directly incorporating users' social preferences as conditions in the diffusion recommendation model, we aim to mitigate popularity bias by strategically adjusting the weights of long-tail and hot items within the social preferences.
Based on social homophily theory~\cite{mcpherson2001birds}, we first explore the user's social preference with matrix multiplication as:
\begin{align}
    \mathbf{R}^{s} = \mathbf{S} \mathbf{R},
\end{align}the element $r_{ij}^{s}$ in $\mathbf{R}^{s}$ denotes the sum of interactions from all social neighbors of user $u_i$ to item $v_j$. The larger value of $r_{ij}^{s}$ indicates that user $u_i$ is more likely to be interested in item $v_j$.
Then, we invert the social preference matrix $\mathbf{R}^s$ to yield a matrix $\mathbf{R}^{sdb}$ designed to mitigate popularity bias,
which can be computed as follows:
\begin{align}
r^{sdb}_{ij} = f(r^s_{ij}),
\end{align}
\begin{align}\label{obtain_xsdb}
f(r^s_{ij}) = 
\begin{cases} 
1 / r^s_{ij} & \text{if } r^s_{ij} \neq 0 \\
0 & \text{otherwise} 
\end{cases},
\end{align}where $f(\cdot)$ is a function designed to adjust user social preferences, aimed at mitigating the popularity bias of the model. $r^{sdb}_{ij}$ and $r^{s}_{ij}$ are the elements located at row $i$ and column $j$ in $\mathbf{R}^{sdb}$ and $\mathbf{R}^s$, respectively.

\subsubsection{Introducing social information as a condition into diffusion recommendation model}
In order to take advantage of the rich information in social networks, we propose a Condition-Guided Diffusion Recommendation Model (CGD) to introduce the user's social preference as a condition in the diffusion recommendation model to improve the recommendation performance and mitigate the popularity bias. 
Specifically, we incorporate the user's social preference matrix $\mathbf{R}^{s}$ into the user-item interaction matrix $\mathbf{R}$ as condition, which can be formulated as:
\begin{align}\label{inter_u-u_to_u-i}
    \mathbf{R}' = \lambda \mathbf{J} \odot f(\mathbf{R}^{s}) + \mathbf{R},
\end{align}where $\mathbf{J} \in\mathbb{R}^{m\times{n}}$ is the matrix whose elements are all $1$. $\odot$ denotes the element-wise product. $\lambda$ is a coefficient to control the effect of the user's social preferences on the model's recommendation results, thus avoiding that the user's social preferences dominate the conditional recommendation results. 
Finally, we incorporate the user's social preference in each inferencing step of the inference process of the diffusion recommendation model as follows:
\begin{align}
    p_{\theta}(\mathbf{x}_{i,t-1}|\mathbf{x}_{i,t}, \mathbf{x}_{i}') = \mathcal{N}(\mathbf{x}_{i,t-1};\mu_{\theta}'(\mathbf{x}_{i,t},\mathbf{x}_{i}',t), \Sigma_{\theta}),
\end{align}
\begin{align}\label{x_rdp}
\mu_{\theta}'(\mathbf{x}_{i,t},\mathbf{x}_{i}',t)=(1-\gamma) \mu_{\theta}(\mathbf{x}_{i,t},t)+\gamma\mu_{\theta}(\mathbf{x}_{i}',t),
\end{align}where $\mu_{\theta}'(\mathbf{x}_{i,t},\mathbf{x}_{i}',t)$ and $\Sigma_{\theta}=\Sigma_{\theta}(\mathbf{x}_{i,t},t)$ are the Gaussian parameters. 
$\mathbf{x}_i'=[x_{i1}', x_{i2}', ..., x_{in}']$ is the row vector in the interaction matrix $\mathbf{R}'$ with the condition added.
Significantly, the hyper-parameter $\gamma$ is employed in our CGD to regulate the degree in the inferencing process integrating the user's social preferences.
Through the implementation of the aforementioned approach, our model effectively refers users' social preferences during the item recommendation process. 

\begin{algorithm}
    \caption{The Joint Inference of CGSoRec}\label{algor:Joint_Inference}
    \textbf{Input}: The user-user interaction vectors $\mathbf{s}_i$ of user $u_i$, the user-item interaction vectors $\mathbf{x}_i$ of user $u_i$, the optimised parameter $\theta'$ and $\theta$. \\
    \textbf{Output}: The interaction vector $\bar{\mathbf{x}}_{i}$ for user $u_i$.

    \begin{algorithmic}[1] 
        \STATE Sample $\epsilon\sim\mathcal{N}(0,I)$.
        \STATE Obtain $\mathbf{s}_{i}'$ via Eq.~\ref{obtain_scpl} and Eq.~\ref{obtain_condition_s}.
        \STATE Obtain $\mathbf{s}_{i,T}$, $\mathbf{s}_{i,T}'$ via Eq.~\ref{fdp}.
        \WHILE{$t=T,\cdots,1$}
        \STATE Calculate $\hat{\mathbf{s}}_{i,t-1}$ and $\hat{\mathbf{s}}_{i,t-1}'$ with $\hat{\mathbf{s}}_{i,t}$, $\mathbf{s}_{i}'$, $\hat{\mathbf{s}}_{i,t}'$ and $\mu_{\theta'}(\cdot)$ via Eq.~\ref{s_rdp}.
        \ENDWHILE
        \STATE Obtain the final denoising result $\bar{\mathbf{s}}_{i}$ via Eq.~\ref{final_s}.
        \STATE Obtain $\mathbf{x}_{i}'$ with via Eq.~\ref{inter_u-u_to_u-i}.
        \STATE Obtain $\mathbf{x}_{i,T}$, $\mathbf{x}_{i,T}'$ via Eq.~\ref{fdp}.
        \WHILE{$t=T,\cdots,1$}
        \STATE Calculate $\hat{\mathbf{x}}_{i,t-1}$ and $\hat{\mathbf{x}}_{i,t-1}'$ with $\hat{\mathbf{x}}_{i,t}$, $\mathbf{x}_{i}'$, $\hat{\mathbf{x}}_{i,t}'$ and $\mu_{\theta}(\cdot)$ via Eq.~\ref{x_rdp}.
        \ENDWHILE
        \STATE Obtain the final recommendation result $\bar{\mathbf{x}}_{i}$ via Eq.~\ref{final_x}.
        \STATE \textbf{return} recommendation result $\bar{\mathbf{x}}_{i}$
    \end{algorithmic}
    
\end{algorithm}

\subsection{Model Training}

In generally, 
the loss function $\mathcal{L}_{CGD}$ of the Condition-Guided Diffusion Recommendation Model (CGD) can be divided into two components~\cite{wang2023diffusion}: the denoising matching loss function $\mathcal{L}_{t}$ and the reconstruction loss function $\mathcal{L}_{1}$, which can be calculated by:
    \begin{align}\label{sgd_loss_funcs}
\mathcal{L}_{1} &=\mathbb{E}_{q(\mathbf{x}_{i,1}|\mathbf{x}_{i,0})}[{\Vert\hat{x}_{\theta}(\mathbf{x}_{i,1},1)-\mathbf{x}_{i,0}\Vert}_{2}^{2}],
         \\
        \mathcal{L}_{t} &= \mathbb{E}_{q(\mathbf{x}_{i,t}|\mathbf{x}_{i,0})}[\frac{1}{2}(\frac{\overline{\alpha}_{t-1}}{1-\overline{\alpha}_{t-1}}-\frac{\overline{\alpha}_{t}}{1-\overline{\alpha}_{t}}){\Vert\hat{x}_{\theta}(\mathbf{x}_{i,t},t)-\mathbf{x}_{i,0}\Vert}_{2}^{2}],  
    \end{align}
where $\hat{x}_{\theta}(\mathbf{x}_{i,t},t)$ is the neural network with learnable parameters $\theta$ predicting $\mathbf{x}_{i,0}$ based on $\mathbf{x}_{i,t}$. We can optimize the parameter $\theta$ in $\hat{x}_{\theta}(\mathbf{x}_{i,t},t)$ by minimizing $\mathcal{L}_{CGD} = \sum_{t=1}^T \mathcal{L}_t$.
The loss function $\mathcal{L}_{CSD}$ of the Condition-Guided Social Denoising Model (CSD) is similar to that of CGD.

\subsection{Joint Inferencing}
To effectively mitigate the popularity bias inherent in recommendation model, we design a joint inference between the Condition-Guided Diffusion Recommendation Model (CGD) and the Condition-Guided Social Denoising Model (CSD) to obtain the final recommendation results.
The overall process of joint inferencing is illustrated in Figure~\ref{joint_inference}. 
Specifically, with CSD, we first acquire the corrupted user social interaction vector $\mathbf{s}_{i, t}$ and corrupted user social interaction vector $\mathbf{s}_{i, t}'$.
Then, we execute a $t$-step denoising operation on the corrupted interaction vector $\mathbf{s}_{i, t}$, $\mathbf{s}_{i, t}'$ and introduce the interaction vector $\mathbf{s}_{i}'$ as condition in each denoising step of $\mathbf{s}_{i, t}$ to guide the denoising result via the neural networks $\mu_{\theta'}$ in Eq.~\ref{s_rdp}.
At last, we balance the relation between $\hat{\mathbf{s}}_{i}'$ and $\hat{\mathbf{s}}_{i}$ by hyper-parameter $w_s$:
    \begin{align} \label{final_s}
    \bar{\mathbf{s}}_{i} = (1-w_s)\hat{\mathbf{s}}_{i} + w_s\hat{\mathbf{s}}_{i}',
\end{align}where $\hat{\mathbf{s}}_{i} = (1-\eta) \mu_{\theta'}(\mathbf{s}_{i,1},1)+\eta\mu_{\theta'}(\mathbf{s}_{i}',1)$. 
We use $\bar{\mathbf{s}}_{i}$ for user ranking and derive the ultimate denoising result of social relation for user $u_i$.
Then, we apply a similar approach as CSD to obtain the corrupted user-item interaction vector $\mathbf{x}_{i,t}$ and  $\mathbf{x}_{i,t}'$ and execute a $t$-step denoising operation on them via the neural networks $\mu_{\theta}$ in Eq.~\ref{x_rdp} to obtain the recommendation results $\hat{\mathbf{x}}_{i}'$ and $\hat{\mathbf{x}}_{i}$.
And we also balance the relation between $\hat{\mathbf{x}}_{i}'$ and $\hat{\mathbf{x}}_{i}$ by hyper-parameter $w_r$, which can be formulated as:
    \begin{align} \label{final_x}
    \bar{\mathbf{x}}_{i} = (1-w_r)\hat{\mathbf{x}}_{i} + w_r\hat{\mathbf{x}}_{i}',
\end{align}where $\hat{\mathbf{x}}_{i} = (1-\gamma) \mu_{\theta}(\mathbf{x}_{i,1},1)+\gamma\mu_{\theta}(\mathbf{x}_{i}',1)$. 
The joint inference procedure of our proposed model is summarized in Algorithm~\ref{algor:Joint_Inference}.

\subsection{Model Analysis}
Let $m$ denote the number of users and $n$ the number of items. In the joint inference phase, CGSoRec utilizes two conditional diffusion models (CSD and CGD) to provide unbiased recommendations to users. At each time step $t$, CSD requires $O(m)$ to denoise the social relationships of users, while CGD requires $O(n)$ to generate unbiased recommendations for each user based on the denoised social relationships.
The time complexity of CGSoRec depends solely on the number of users, items, and time steps $T$.
Therefore, the overall time complexity of CGSoRec during the inference phase is $O(T(m+n))$.
In contrast, traditional recommendation models typically have a time complexity of $O(mn)$ with $m,n \gg T$ in most cases, because they require the computation of similarity between each user and all items.
This implies that CGSoRec is more scalable to larger datasets compared to traditional methods.

\section{Experiments}

In this section, we perform comprehensive experiments on three real-world datasets over three classic social recommendation models and our proposed CGSoRec to answer the following questions (RQs):
\begin{itemize}
    \item \textbf{(RQ1)} How does the integration of social networks impact the performance and popularity bias of the \textbf{CGSoRec} compared to existing recommendation methods?
    \item \textbf{(RQ2)} How significant is the performance improvement for both hot and long-tail items when \textbf{CGSoRec} compared with \textbf{DiffRec} and \textbf{MACR}?
    \item \textbf{(RQ3)} What are the contributions of different components of \textbf{CGSoRec} to performance?
    \item \textbf{(RQ4)} How do the hyper-parameters affect the performance of the proposed method \textbf{CGSoRec}?
\end{itemize}

\subsection{Experimental Settings}

\subsubsection{Datasets} To evaluate the effectiveness of the proposed method, we performed comprehensive experiments on three representative recommendation datasets with social information, including LastFM~\cite{cantador2010content}, DBook~\cite{lu2020meta}, and Ciao~\cite{tang2012mtrust}. The statistical summary of these three datasets is shown in Table~\ref{tab:dataset}.

\begin{itemize}
    \item \textbf{LastFM}~\cite{cantador2010content}. This dataset is sourced from last.fm\footnote{\url{https://www.last.fm}}, a UK-based Internet radio and music community. It is particularly valuable for testing social recommendation models due to its rich social network data. Each user in the dataset is associated with a list of their favorite artists, along with the corresponding play counts. 
    Additionally, the dataset provides user application tags, which can be leveraged to create content vectors. 
    In total, this dataset comprises 1,853 users and 2,698 items.    
    \item \textbf{DBook}~\cite{lu2020meta}. This dataset is sourced from Douban.Book\footnote{\url{https://book.douban.com}}, a recommendation system dataset compiled from the Douban website. 
    It includes user ratings of books on a scale from 1 to 5, along with data on social connections among users. 
    This dataset is the largest dataset of the three datasets considered in this paper, featuring 10,864 users and a collection of 22,346 distinct items.
    \item \textbf{Ciao}~\cite{tang2012mtrust}. This dataset is provided by a product review website called Ciao\footnote{\url{http://www.ciao.com}}.
    It includes user ratings for items they have purchased and social connections between users. Since we focus on implicit feedback, we convert detailed ratings into binary values of 0 or 1, indicating whether the user is interested in the item or not.
    The version of this dataset used in our proposed method contains 5,868 users and 10,724 items.
    \item Gowalla~\cite{he2020lightgcn}: This dataset is collected from LightGCN\footnote{https://github.com/gusye1234/LightGCN-PyTorch}, where users share their locations through check-ins. To maintain data quality, we retains only users and items with at least ten interactions. To incorporate social networks, we simulate a social network based on co-purchase relationships in this dataset, where each user retains a subset of users with frequent co-purchases as their friends.
\end{itemize}

\begin{table}[h]
\centering
\renewcommand\arraystretch{1}
\caption{Statistics of the experimental data.}
\label{tab:dataset}
\scalebox{1}{
\begin{tabular}{|c|c|c|c|c|c|}
    \specialrule{0.08em}{0pt}{0pt}
    Dataset & LastFM & DBook & Ciao & Gowalla\\
    \hline \hline
    \#of users& 1,853 & 10,864 & 5,868 &29,858\\ \hline
    \#of items& 2,698 & 22,346 & 10,724&40,981\\  \hline
    \#of Interactions ($\mathbf{R}$) & 46,542 & 787,286 & 143,217&1,027,370\\ \hline

    \#of Density ($\mathbf{R}$)& 0.00931 & 0.00324 & 0.00228&0.00084\\ \hline
    \#of Connections ($\mathbf{S}$)& 25,435 & 169,151 & 111,782&/\\ \hline
    \#of Density ($\mathbf{S}$)& 0.00741 & 0.00143 & 0.00335&/\\ \hline
    \specialrule{0.08em}{0pt}{0pt}
\end{tabular}}
\end{table}

\subsubsection{Evaluation metrics} 
We employ two widely recognized metrics to assess the performance of top-K recommendations. Here, $K$ represents the number of recommendations provided to users by the model. While a larger $K$ can yield more stable evaluation results, it may not accurately reflect the targeted quality of the recommendations from a user's perspective. In our experiments, we set $K$ to 5 and 10. Higher values of these metrics indicate superior performance of the model in making recommendations. The mathematical definitions of these evaluation metrics are as follows: 
\begin{itemize}
\item \textbf{Recall@K}. This metric assesses the model's capability to recommend a list of items to users:
\begin{equation}
    Recall@K = \frac{\sum_{u_i\in U}|R(u_i)\cap T(u_i)|}{\sum_{u_i\in U}|T(u_i)|}
\end{equation}
where $R(u_i)$ represents the list of recommendations made to the user, derived from their interactions in the training set, $T(u_i)$ corresponds to the user's interactions in the testing set.

\item \textbf{NDCG@K}. This metric evaluates the quality of the recommendations by considering positional importance of each item in the recommended list:
\begin{align}
    DCG_{u_i}\textbf{@}K = \sum_{(u_i,v_j)\in D^{\text{test}}}\frac{\textbf{I}(\hat{Z}_{u_i,v_j}\leq K)}{\log(\hat{Z}_{u_i,v_j} + 1)}
\end{align}
\begin{align}
    NDCG\textbf{@}K = \frac{1}{|U|}\sum_{u_i\in U}\frac{DCG_{u_i}\textbf{@}K}{IDCG_{u_i}\textbf{@}K}
\end{align}
where $IDCG_{u_i}\textbf{@}K$  is the ideal $DCG_{u_i}\textbf{@}K$. $\hat{Z}_{u_i,v_j}$ denotes the rank position of a positive feedback $(u_i,v_j)$. $D^{\text{test}}$ is the testing set.
\end{itemize}





\subsubsection{Parameter Setting}
In the training process of our proposed method, we set the parameters as: the batch size is set to 400 as with DiffRec. The hidden layer size of the model $d$ is searched in the range $\{[300], [200, 600], [1000], [1500]\}$.
We choose the learning rates $lr$ in $\{0.0001, 0.0005, 0.001, 0.005\}$ of our proposed method, and perform a grid search to find the best parameters. 
We divide the interactions into train, test, and valid set with a ratio of $8:1:1$.
The specific experiment is described below. We use Adam to optimize the model's parameters. 
For a fair comparison, we closely follow the default setting of the baselines.

\subsubsection{Dataset splitting}
It is important to note that when dealing with randomly sliced recommendation datasets, both training and test sets exhibit long-tail distributions. This can lead to improved results even if the model's popularity bias is intensified, consequently masking the adverse effects of popularity bias on the recommendation model. To address this, we draw on previous works~\cite{wei2021model,quan2023robust} to construct a debiased testing set where the interactions are sampled to be a uniform distribution over items, providing a more accurate measure of model performance. 
Specially, the better the model performs on the debiased test set, the lower the popularity bias, and vice versa. This allows us to compare popularity bias across models using Recall and NDCG on the debiased test set.

\subsubsection{Baseline models}
To evaluate the effectiveness of \textbf{CGSoRec}, we compare the proposed method with representative social recommendation models, including the graph-based social recommendation models (\textbf{DiffNet}~\cite{wu2019neural}, \textbf{LightGCN-S}~\cite{he2020lightgcn}), the self-supervised learning-based social recommendation model (\textbf{MHCN}~\cite{yu2021self}), diffusion recommendation model(\textbf{DiffRec}~\cite{wang2023diffusion}) and the debiased recommendation model( \textbf{MACR}~\cite{wei2021model}).
These baseline methods are summarized as follows:

\begin{itemize}
\item DiffNet~\cite{wu2019neural}: This is a GCN-based social recommendation method that models the recursive dynamic social diffusion in both user and item spaces. 
We omit the multiplication operation between the user feature matrix and the user-user interaction matrix at each layer of the model to eliminate the influence of the social network.
\item LightGCN-S: This is an enhanced model, building upon LightGCN~\cite{he2020lightgcn}, that integrates social network information to boost the performance of recommendation systems. 
We exclude the user-user interaction matrix from the adjacency matrix to remove the influence of the social network.
\item MHCN~\cite{yu2021self}: This is a self-supervised learning-based social recommendation method that integrates hypergraph convolutional networks. 
We omit the social self-supervised learning loss within the target loss function to eliminate the impact of social networks.
\item GDMSR~\cite{quan2023robust}: This is a social recommendation method which improves graph-based social recommendation by leveraging user-item feedback to filter noisy social relations and enhance preference learning, addressing the lack of reliably labeled data. We incorporate this method into MHCN to validate its effectiveness.
\item DiffRec-S: This is an enhanced model based on DiffRec~\cite{wang2023diffusion} that learns the generative process in a denoising manner to reconstruct the user-item interaction matrix. We incorporate unadjusted user social preferences as conditions into the training process of the diffusion model.
\item MACR-S: This is a social network enhancement method based on MACR~\cite{wei2021model} that mitigates the popularity bias of recommendation models by causal learning. We incorporate user social networks into the adjacency matrix $\textbf{A}$, thereby integrating the social networks into the MACR model. 

\end{itemize}

\begin{table*}[!t]
\centering
\caption{Performance changes comparison of different social recommendation methods under a debiased testing set.}
\label{tab:performance_com}
\scalebox{0.52}{
\begin{tabular}{clcccccccccccc}
\toprule
\multicolumn{2}{c}{\textbf{Datasets}}              & 
\multicolumn{3}{c}{\textbf{LastFM}} & \multicolumn{3}{c}{\textbf{DBook}}& \multicolumn{3}{c}{\textbf{Ciao}}& \multicolumn{3}{c}{\textbf{Gowalla}}\\ 
\multicolumn{1}{c}{\textbf{Basemodel}}              & \multicolumn{1}{c}{\textbf{Method}} &
\multicolumn{1}{c}{\textbf{R@5}} & \multicolumn{1}{c}{\textbf{R@10}} &\multicolumn{1}{c}{\textbf{N@10}} &
\multicolumn{1}{c}{\textbf{R@5}} & \multicolumn{1}{c}{\textbf{R@10}} &\multicolumn{1}{c}{\textbf{N@10}}&
\multicolumn{1}{c}{\textbf{R@5}} & \multicolumn{1}{c}{\textbf{R@10}} &\multicolumn{1}{c}{\textbf{N@10}}&
\multicolumn{1}{c}{\textbf{R@5}} & \multicolumn{1}{c}{\textbf{R@10}} &\multicolumn{1}{c}{\textbf{N@10}}\\
\midrule
\multirow{3}{*}{\textbf{DiffNet}}
& \textbf{w/o S} &   0.0178    &   0.0323   &  0.0343     &   0.0049     &    0.0077       &0.0091&       0.0026       &    0.0053 & 0.0048 &0.0050&0.0102&0.0256\\  
& \textbf{Ori.} &       0.0165  &0.0313    &     0.0321 &   0.0045    &   0.0072       &  0.0084 &    0.0024         &  0.0049 & 0.0044 &0.0046&0.0094&0.0250\\
& \textbf{Improv.}     &     -7.30\%
    &-3.10\%    &     -6.41\%    &  -8.16\%    &     -6.49\%     &      -7.69\%& -7.69\%       &  -7.54\% & -8.33\%  &-8.00\%&-0.78\%&-2.34\%\\ 
    \midrule
\multirow{3}{*}{\textbf{LightGCN-S}}
& \textbf{w/o S}    &      0.0156    &   0.0334   &  0.0349     &   0.0044    &   0.0071      &  0.0079 &       0.0027 &  0.0054   & 0.0047 &0.0047&0.0099&0.0258\\  
& \textbf{Ori.} &       0.0151  &0.0323    &     0.0344 &      0.0041 &    0.0069      & 0.0075 &   0.0025    & 0.0049 &  0.0044 &0.0042&0.0091&0.0243\\
                                 & \textbf{Improv.}     &    -3.21\%
    &-3.29\%    &     -1.43\%    &  -6.82\%   &    -2.82\%     &   -5.06\%   &   -7.41\%     &  -9.26\% &-6.38\% &-10.64\%&-8.08\%&-5.81\%\\ 
    \midrule
\multirow{3}{*}{\textbf{MHCN}}
& \textbf{w/o S}    &      0.0149 &   0.0328   &  0.0328     &   0.0034   &   0.0060  & 0.0066  &      0.0024       &   0.0042  & 0.0042 &0.0053&0.0113&0.0253\\  
& \textbf{Ori.} &       0.0148  & 0.0304    &     0.0301 &  0.0032    &      0.0055    & 0.0061 &     0.0022        & 0.0037   &0.0038  &0.0051&0.0107&0.0248\\
                                 & \textbf{Improv.}     &     -0.67\%
    &-7.32\%    &     -8.23\%    & -5.88\%   &  -8.33\%  &  -7.58\% &  -8.33\% & -11.90\% & -9.52\% &-3.77\%&-5.31\%&-1.98\%\\ 
\midrule 
& \textbf{w/o S}    &      0.0149 &   0.0328   &  0.0328     &   0.0034   &   0.0060  & 0.0066  &      0.0024       &   0.0042  & 0.0042 &0.0053&0.0113&0.0253\\ 
\textbf{GDMSR}&\textbf{Ori.} &  0.0146&0.0306&0.0296&0.0029&0.0052&0.0060&0.0023&0.0039&0.0040&0.0050&0.0109&0.0251\\  
                                 &\textbf{Improv.}     &-2.01\%&-6.71\%&-9.76\%&-14.71\%&-13.33\%&-9.09\%&-4.17\%&-7.14\%&-4.76\%&-5.66\%&-3.54\%&-0.79\%\\ 
            \midrule
    \multirow{3}{*}{\textbf{MACR-S}}
& \textbf{w/o S}    &   0.0241    &   0.0572   &  0.0454     &   0.0031   &   0.0052  &  0.0097 &  0.0034           &   0.0060  & 0.0052 &0.0093&0.0192&0.0320\\  
& \textbf{Ori.} &      0.0184   &  0.0488   &   0.0379   &    0.0030  &    0.0050   &  0.0094 &    0.0031      &   0.0057 & 0.0051 &0.0081&0.0180&0.0295\\
                                 & \textbf{Improv.}     &     -23.65\%
    & -14.69\%    &    -16.52\%    & -3.23\%   & -3.85\%  &  -3.09\% & -8.82\% & -5.00\% & -1.92\% &-5.38\%&-6.25\%&-7.81\%\\ 
            \midrule
    \multirow{3}{*}{\textbf{DiffRec-S}}
& \textbf{w/o S}   &  0.0188  &   0.0405  &  0.0399    &   0.0061    &  0.0114 &  0.0139  &   0.0026   &   0.0052 & 0.0052  &0.0051&0.0115&0.0263\\  
& \textbf{Ori.} &    0.0171     &   0.0380  &  0.0365    &  0.0049    &    0.0090   & 0.0113 &     0.0025     & 0.0045   & 0.0046 &0.0045&0.0097&0.0252\\
                                 & \textbf{Improv.}     &     -8.51\%
    &-6.17\%    &     -8.52\%    & -19.67\%   & -21.05\%  &  -18.71\% & -3.85\% & -13.46\% & -11.54\% &-11.76\%&-15.65\%&-4.18\%\\ 
    \midrule
    \multirow{3}{*}{\textbf{CGSoRec}}
& \textbf{w/o S}    &  0.0188  &   0.0405  &  0.0399    &   0.0061    &  0.0114 &  0.0139  &   0.0026   &   0.0052 & 0.0052 &0.0051&0.0115&0.0263\\  
& \textbf{Ori.} &    0.0227 &  0.0491    &   0.0442 &  0.0065  &    0.0121   &  0.0150&    0.0031   &  0.0057 &0.0057 &0.0059&0.0131&0.0296\\
                                 & \textbf{Improv.}     &     20.74\%
    &21.23\%    &   10.78\%    &  6.56\%   &     6.14\%    &      7.91\% & 19.23\%     &  9.62\% & 9.62\% &15.69\%&13.91\%&12.55\%\\ 
    \bottomrule
\end{tabular}}
\end{table*}

\begin{table*}[t]
\centering
\caption{Performance on hot and long tail item group under a debiased testing set.}
\label{tab:long_and_tail_performance_com}
\scalebox{0.6}{
\begin{tabular}{clcccccccccc}
\toprule
\multicolumn{3}{c}{\textbf{Datasets}}              & 
\multicolumn{3}{c}{\textbf{LastFM}} & \multicolumn{3}{c}{\textbf{DBook}}& \multicolumn{3}{c}{\textbf{Ciao}}\\ 
\multicolumn{1}{c}{\textbf{Basemodel}}              & \multicolumn{1}{c}{\textbf{Group}} & \multicolumn{1}{c}{\textbf{Method}} &
\multicolumn{1}{c}{\textbf{R@5}} & \multicolumn{1}{c}{\textbf{R@10}} &\multicolumn{1}{c}{\textbf{N@10}} &
\multicolumn{1}{c}{\textbf{R@5}} & \multicolumn{1}{c}{\textbf{R@10}} &\multicolumn{1}{c}{\textbf{N@10}}&
\multicolumn{1}{c}{\textbf{R@5}} & \multicolumn{1}{c}{\textbf{R@10}} &\multicolumn{1}{c}{\textbf{N@10}}\\
\midrule
    \multirow{6}{*}{\textbf{MACR-S}}
& \multirow{3}{*}{\textbf{Hot}} & \textbf{w/o S}   & 0.1046  &  0.1517   &   0.1197  &   0.0259    & 0.0496  & 0.0283 &  0.0074   &  0.0182 & 0.0105  \\  
&& \textbf{Ori.} &  0.1053 &   0.1619  &    0.1254 &  0.0301     & 0.0570  & 0.0328  &  0.0087   & 0.0191  & 0.0111 \\
                                 && \textbf{Improv.}     &     0.67\%
    & 6.72\%    &   4.76\%    &16.22\%   & 14.92\%  &  15.90\% & 17.57\% & 4.95\% & 5.71\% \\ 
    & \multirow{3}{*}{\textbf{Tail}}& \textbf{w/o S}   & 0.0234  &  0.0517   &  0.0390  &     0.0026  & 0.0048  & 0.0091  &   0.0027  &  0.0058 & 0.0049  \\  
&& \textbf{Ori.} &  0.0180 &   0.0462  &   0.0322  &  0.0023    & 0.0044  & 0.0086  &   0.0025  & 0.0054  & 0.0046 \\
                                 && \textbf{Improv.}     &     -23.08\%
    &-10.64\%    &   -17.44\%    &-11.54\%   & -8.33\%  &  -5.49\% & -7.41\% & -6.90\% & -6.12\% \\  
            \midrule
    \multirow{6}{*}{\textbf{DiffRec-S}} & \multirow{3}{*}{\textbf{Hot}}
& \textbf{w/o S}   &  0.1178 &  0.1936   &   0.1372    &   0.0703    &  0.1094 &  0.0761 &   0.0175  &  0.0318 & 0.0209  \\  
&& \textbf{Ori.} &  0.0910 &  0.1827   &  0.1258   &      0.0611  & 0.0910  & 0.0646  &    0.0150  & 0.0299  & 0.0194 \\
                                 && \textbf{Improv.}     &     -22.75\%
    &-6.93\%    &   -8.31\%    & -13.09\%   & -16.82\%  & -15.11\% & -14.29\% & -5.97\% & -7.18\% \\ 
    & \multirow{3}{*}{\textbf{Tail}}
& \textbf{w/o S}   &    0.0167  &   0.0376  &   0.0323     &    0.0035   & 0.0069  & 0.0084  &  0.0017   &  0.0032 &  0.0028 \\  
&& \textbf{Ori.} &  0.0146 &  0.0363   &  0.0303   &     0.0028  & 0.0054  & 0.0069  &   0.0014  & 0.0028 & 0.0024 \\
                                 && \textbf{Improv.}     &     -12.57\%
    &-3.46\%    &   -6.19\%    &-20.00\%   & -21.74\%  &  -17.86\% & -17.65\% & -12.5\% & -14.29\% \\ 
    \midrule
    \multirow{6}{*}{\textbf{CGSoRec}} & \multirow{3}{*}{\textbf{Hot}}
& \textbf{w/o S}   &   0.1178 &  0.1936   &   0.1372 &    0.0703    &  0.1094 &  0.0761  &     0.0175  &  0.0318 & 0.0209   \\  
&& \textbf{Ori.} & 0.0940  &   0.1865  &    0.1227 &     0.0666  &  0.1035 & 0.0748  &   0.0166  & 0.0308  & 0.0202 \\
                                 && \textbf{Improv.}     &     -20.20\%
    & -3.67\%    &   -10.57\%    &-5.26\%   & -5.39\%  &  -1.71\% & -5.14\% & -3.14\% & -3.35\% \\  
    &\multirow{3}{*}{\textbf{Tail}}
& \textbf{w/o S}   & 0.0167  &   0.0376  &   0.0323  &      0.0035   & 0.0069  & 0.0084   &    0.0017   &  0.0032 &  0.0028   \\  
&& \textbf{Ori.} &  0.0210 &  0.0461   &   0.0366  &     0.0038  & 0.0076  & 0.0094  &   0.0023  &  0.0037 & 0.0032 \\
                                 && \textbf{Improv.}     &     25.75\%
    & 22.61\%    &   13.31\%    &8.57\%   & 10.14\%  &  11.90\% & 35.29\% & 15.63\% & 14.29\% \\  
    \bottomrule
\end{tabular}}
\end{table*}

\subsection{Performance and popularity bias changes comparison (RQ1)}
\begin{figure*}[h]
\centering
{\subfigure
{\includegraphics[width=0.19\linewidth]{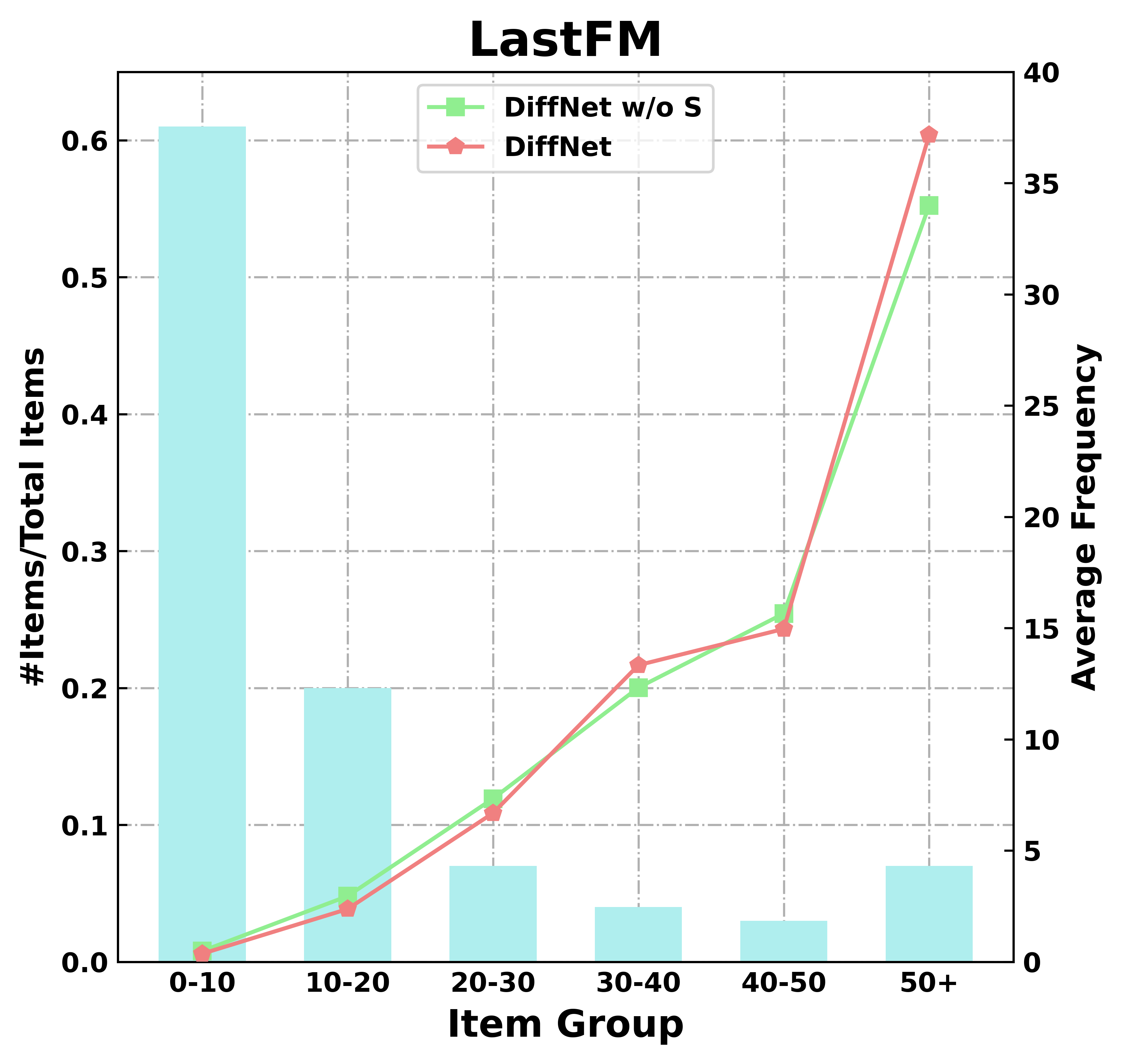}}}
{\subfigure
{\includegraphics[width=0.19\linewidth]{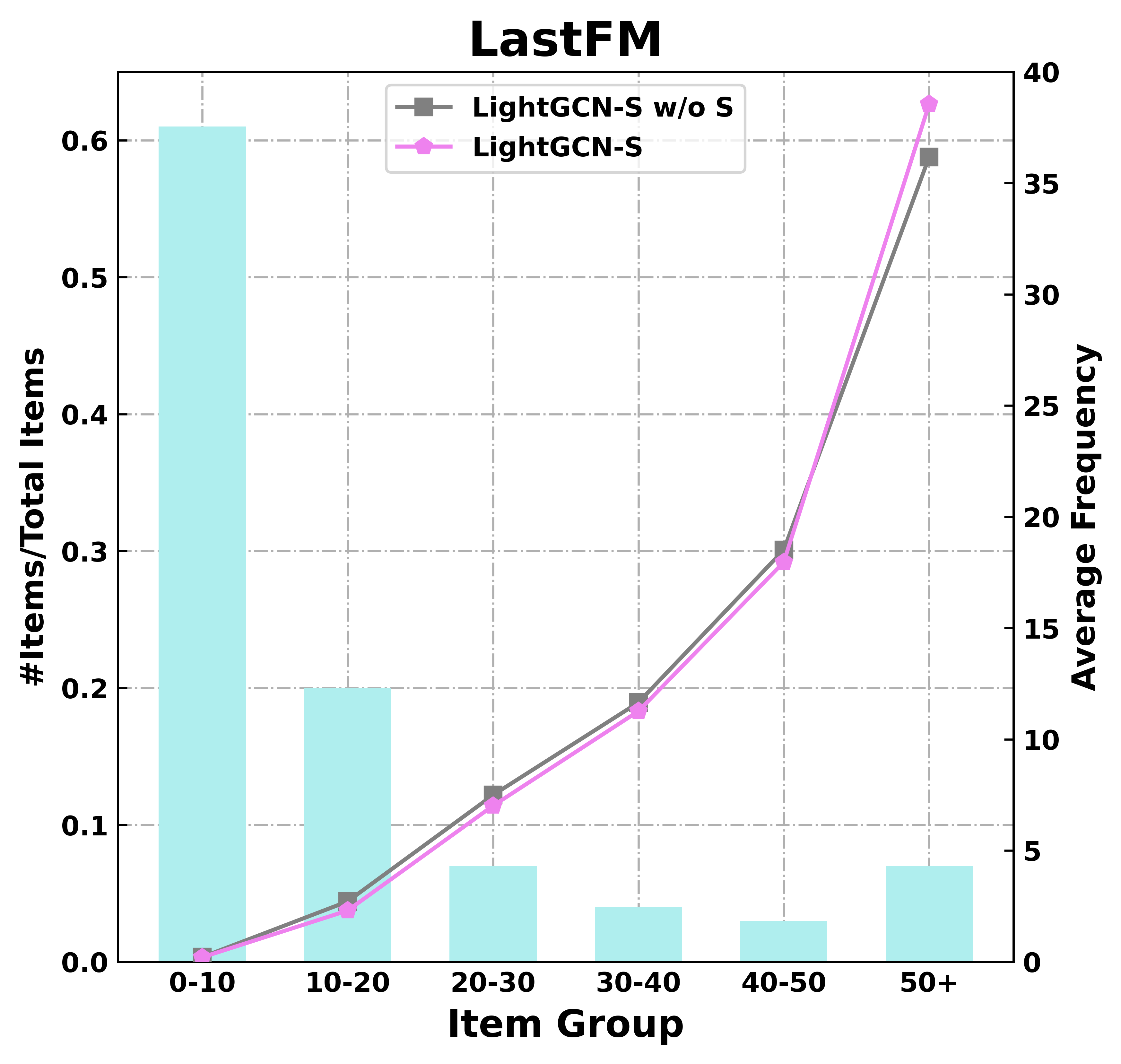}}}
{\subfigure
{\includegraphics[width=0.19\linewidth]{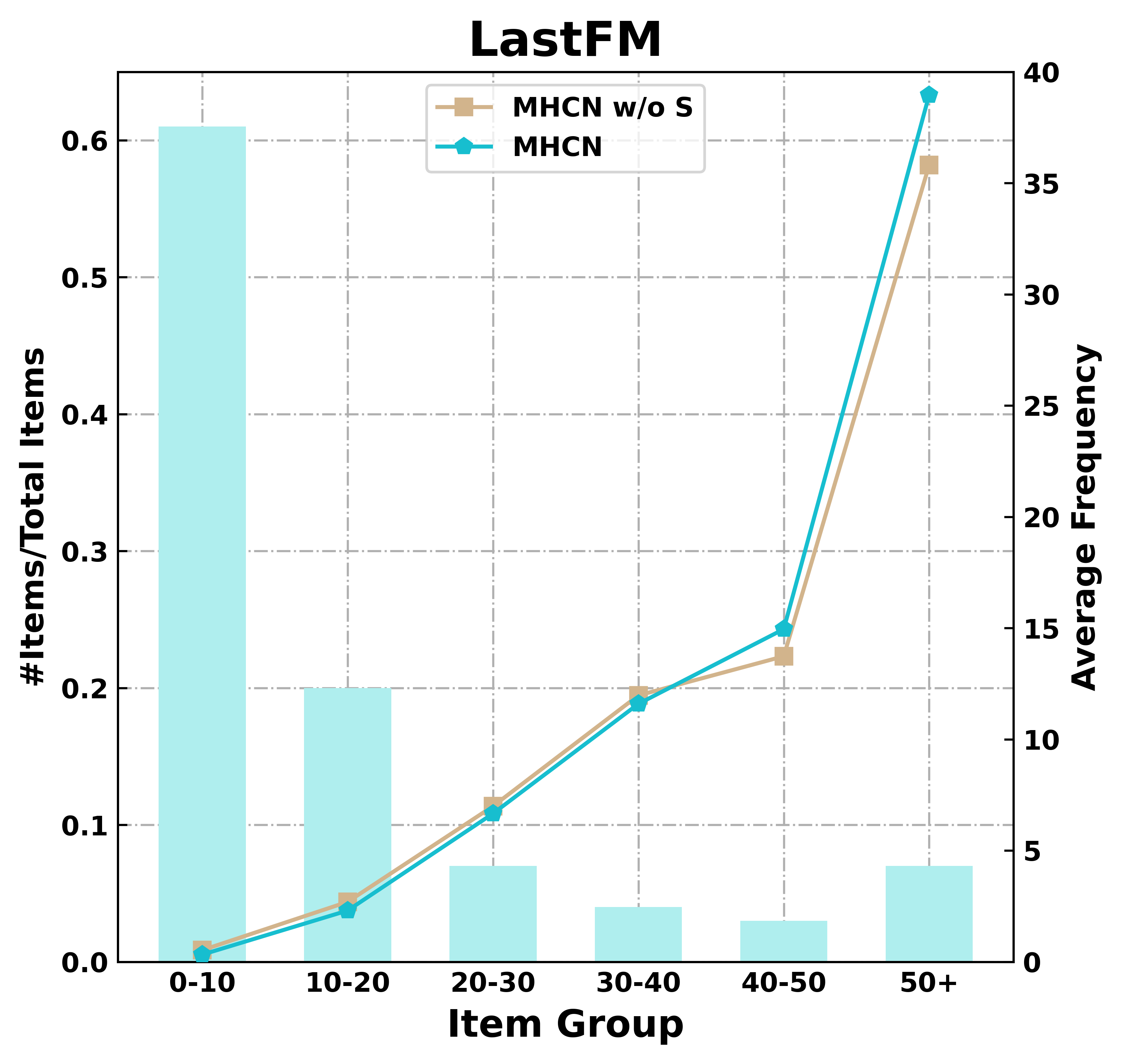}}}
{\subfigure
{\includegraphics[width=0.19\linewidth]{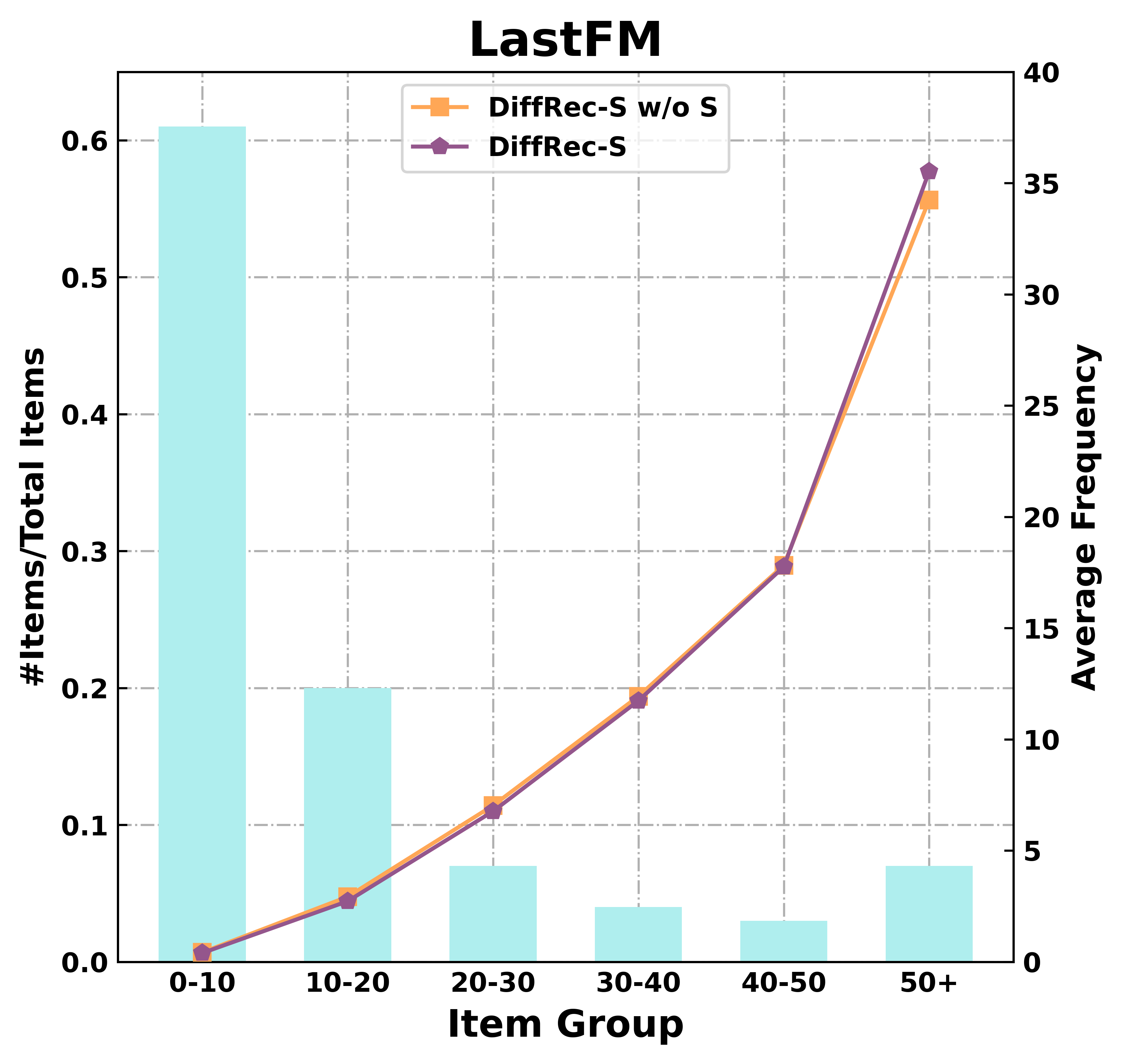}}}
{\subfigure
{\includegraphics[width=0.19\linewidth]{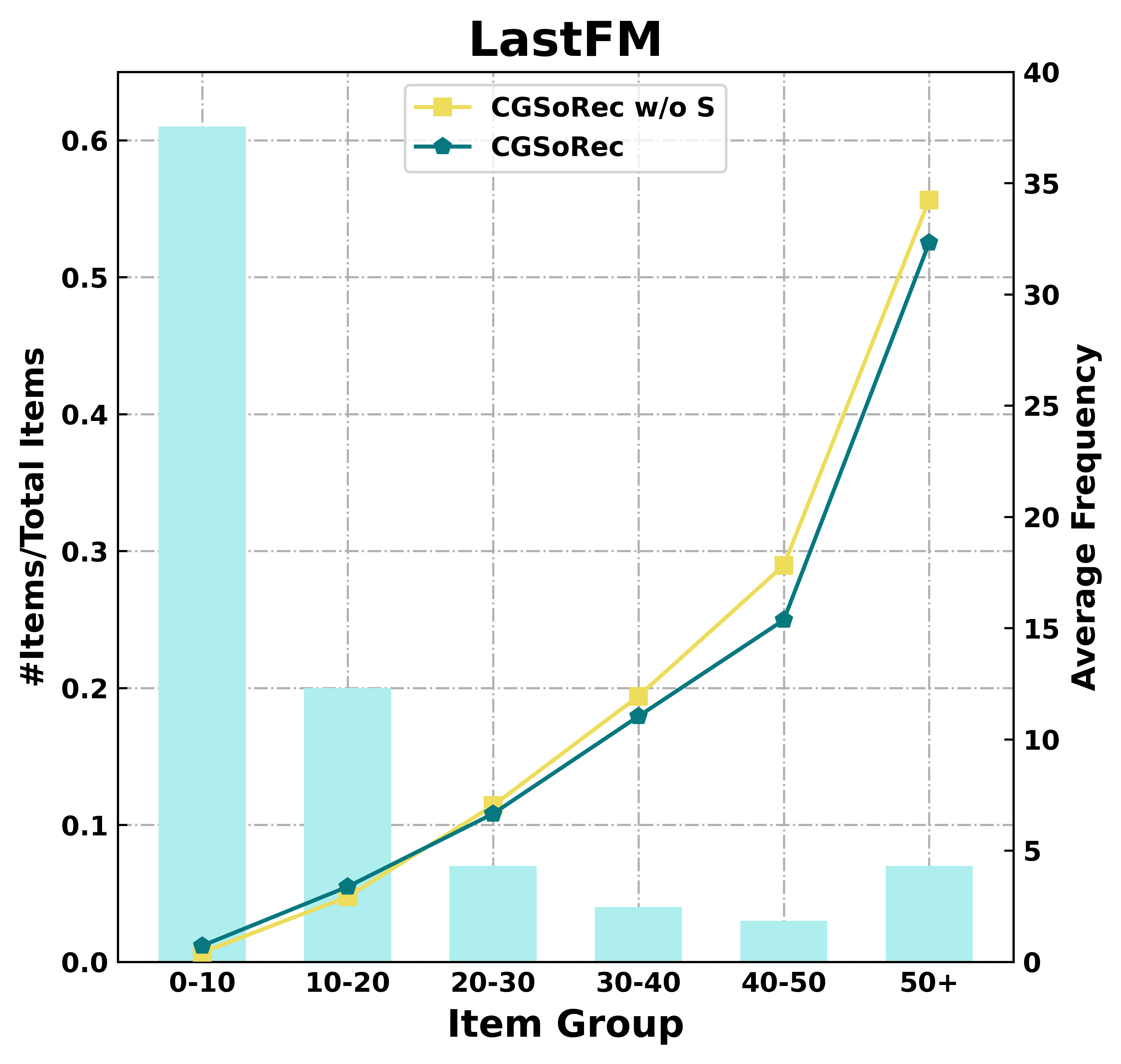}}}%

{\subfigure
{\includegraphics[width=0.19\linewidth]{{ DBook_DiffNet_s_ws.png}}}}
{\subfigure
{\includegraphics[width=0.19\linewidth]{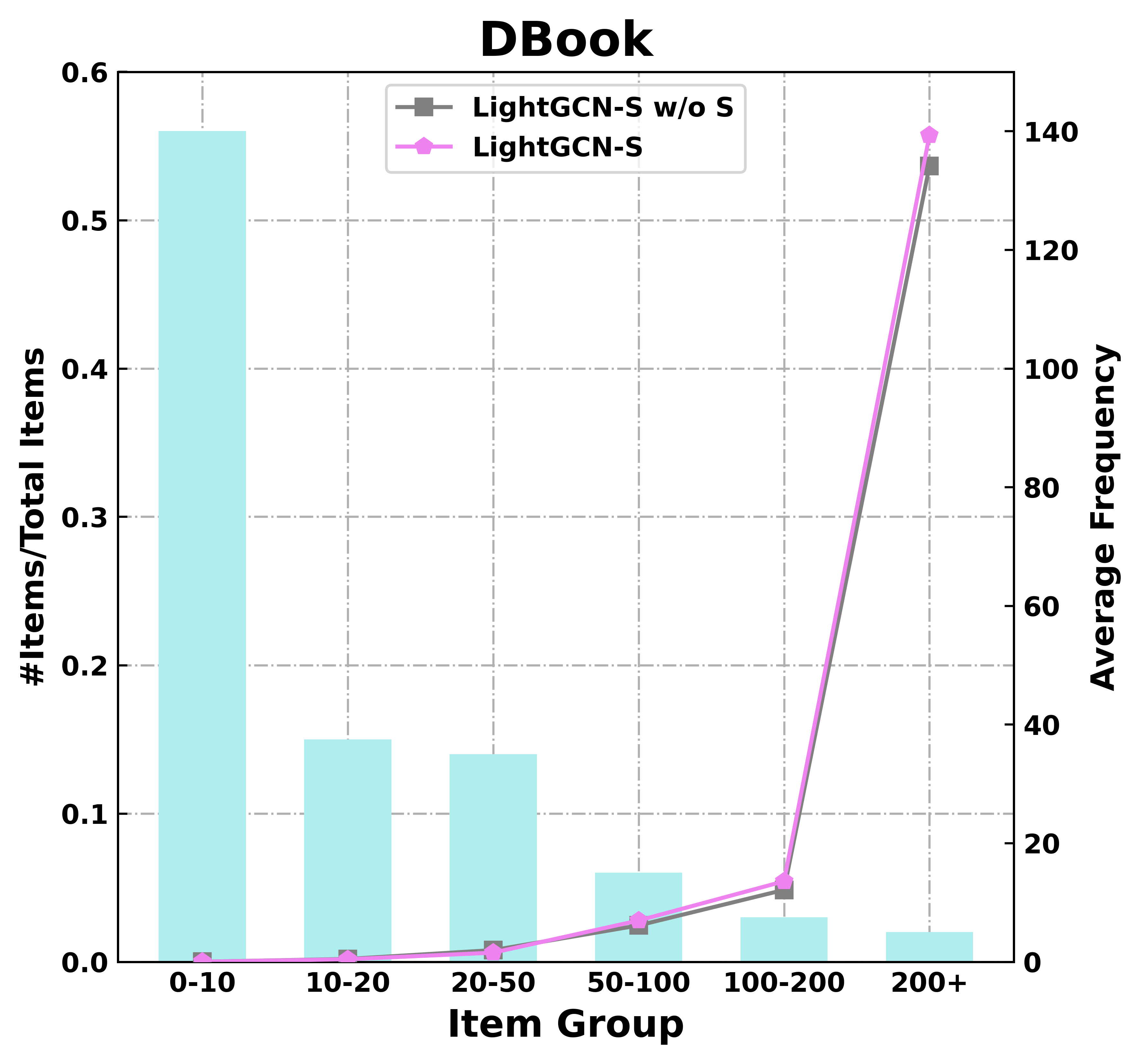}}}
{\subfigure
{\includegraphics[width=0.19\linewidth]{{ DBook_MHCN_s_ws.png}}}}
{\subfigure
{\includegraphics[width=0.19\linewidth]{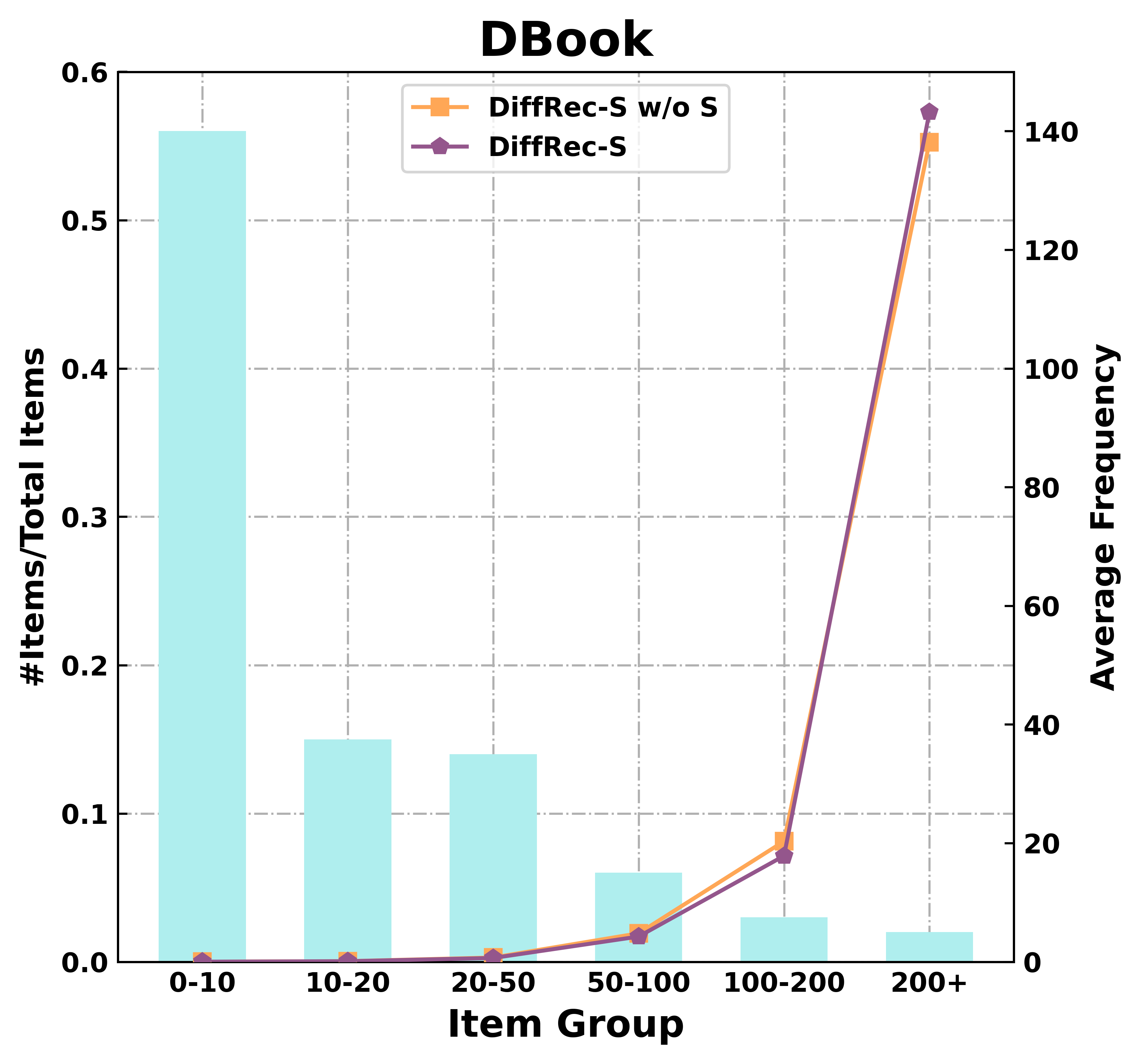}}}
{\subfigure
{\includegraphics[width=0.19\linewidth]{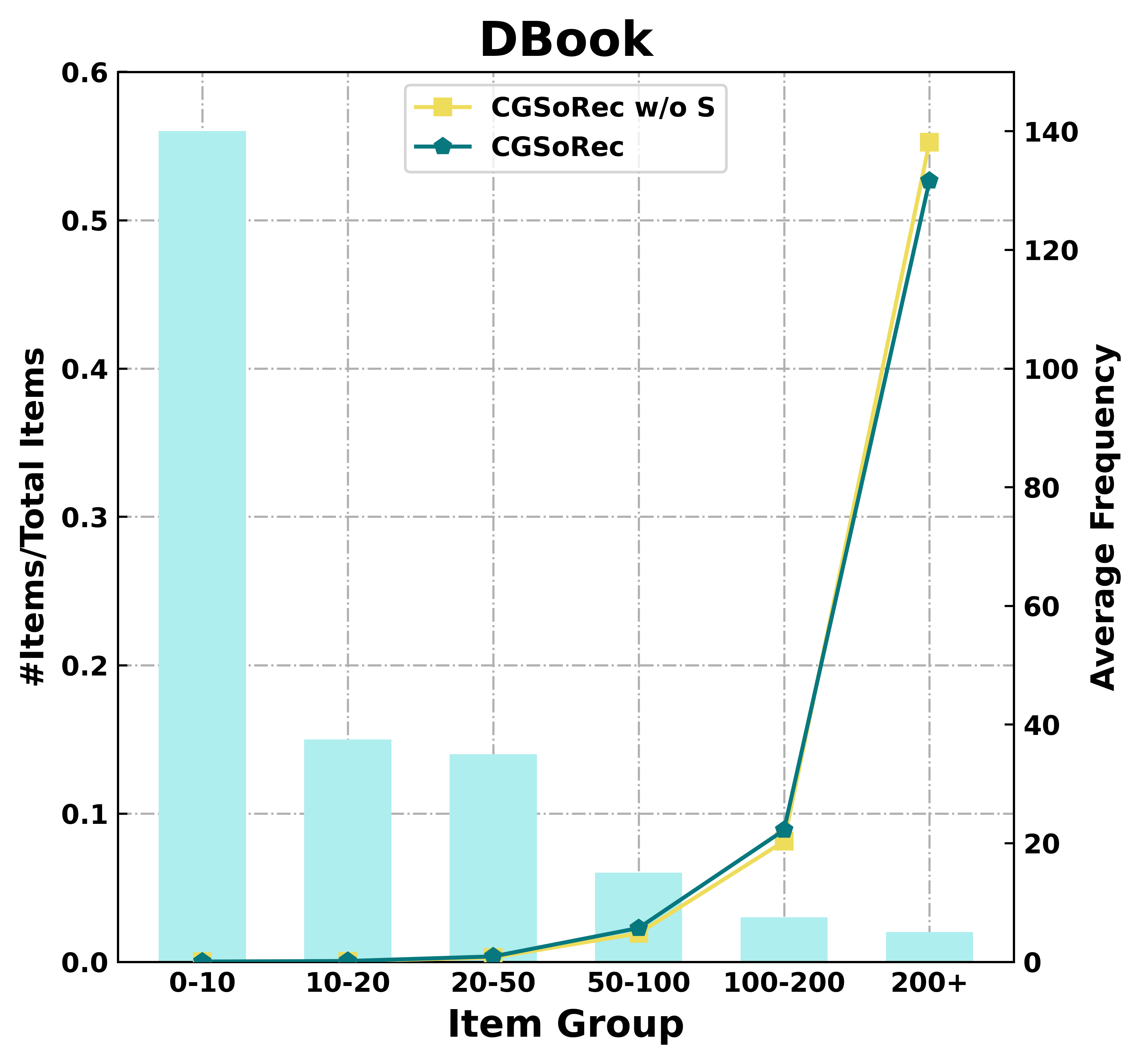}}}%

{\subfigure
{\includegraphics[width=0.19\linewidth]{{ Ciao_DiffNet_s_ws.png}}}}
{\subfigure
{\includegraphics[width=0.19\linewidth]{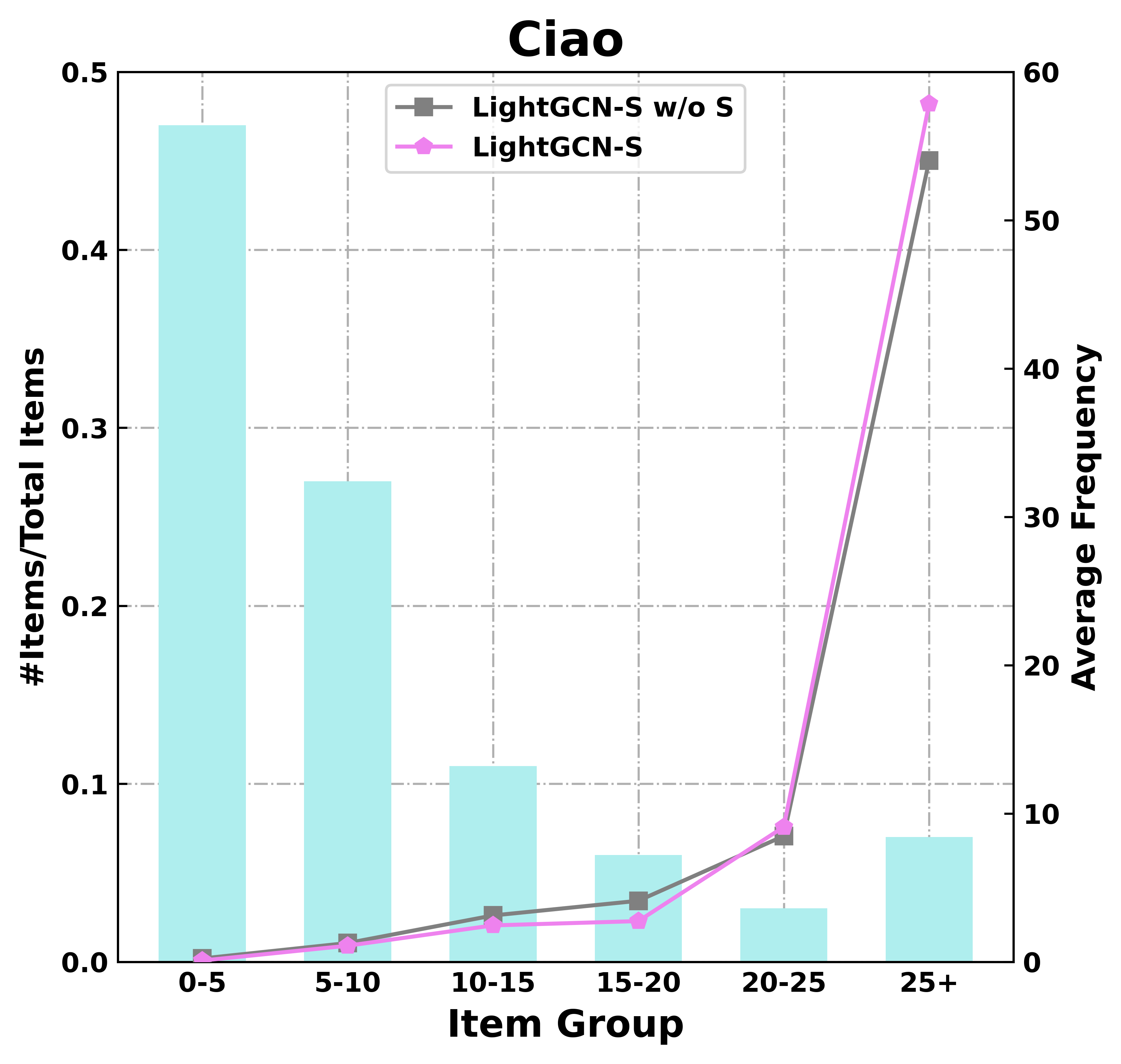}}}
{\subfigure
{\includegraphics[width=0.19\linewidth]{{ Ciao_MHCN_s_ws.png}}}}
{\subfigure
{\includegraphics[width=0.19\linewidth]{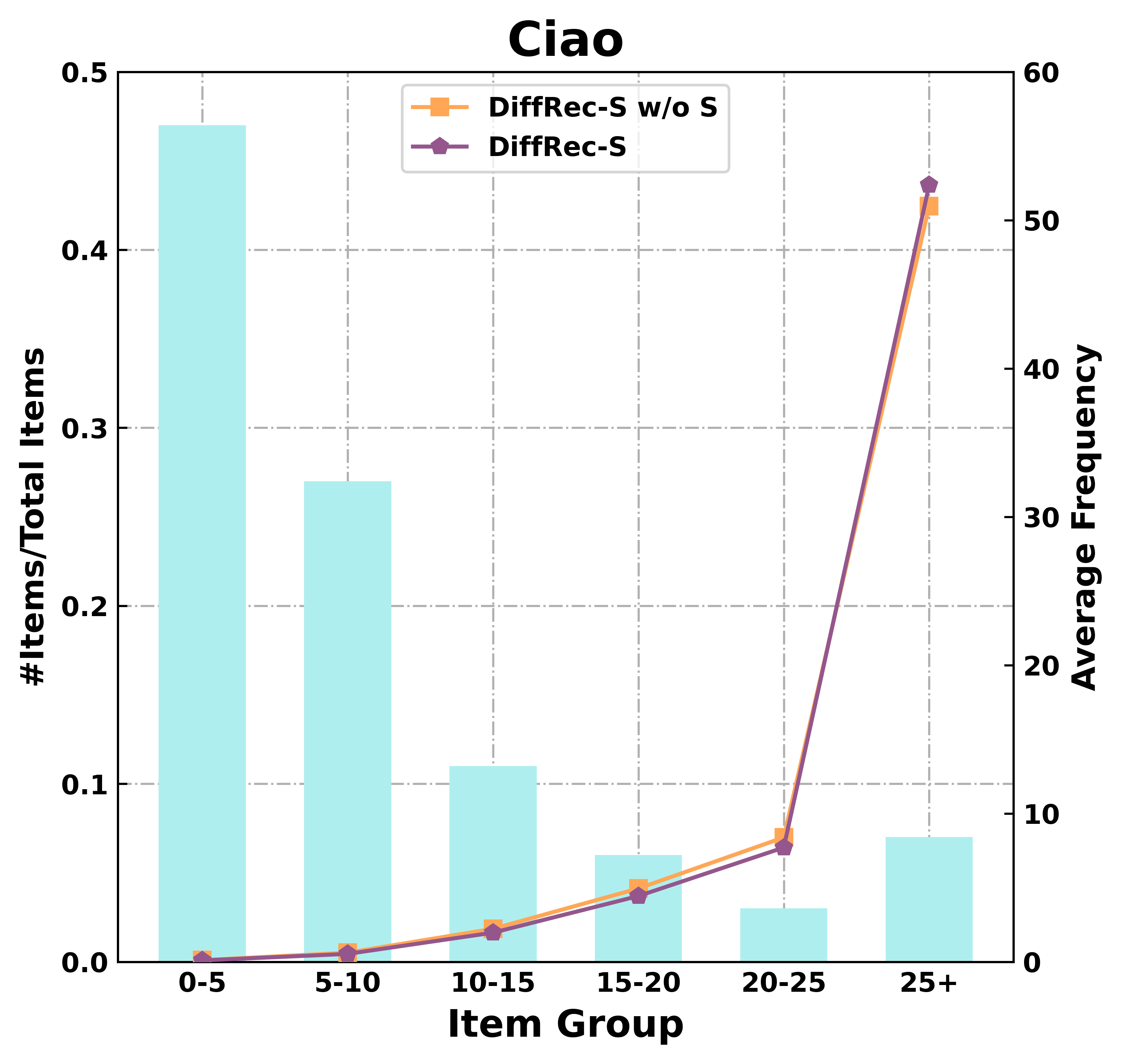}}}
{\subfigure
{\includegraphics[width=0.19\linewidth]{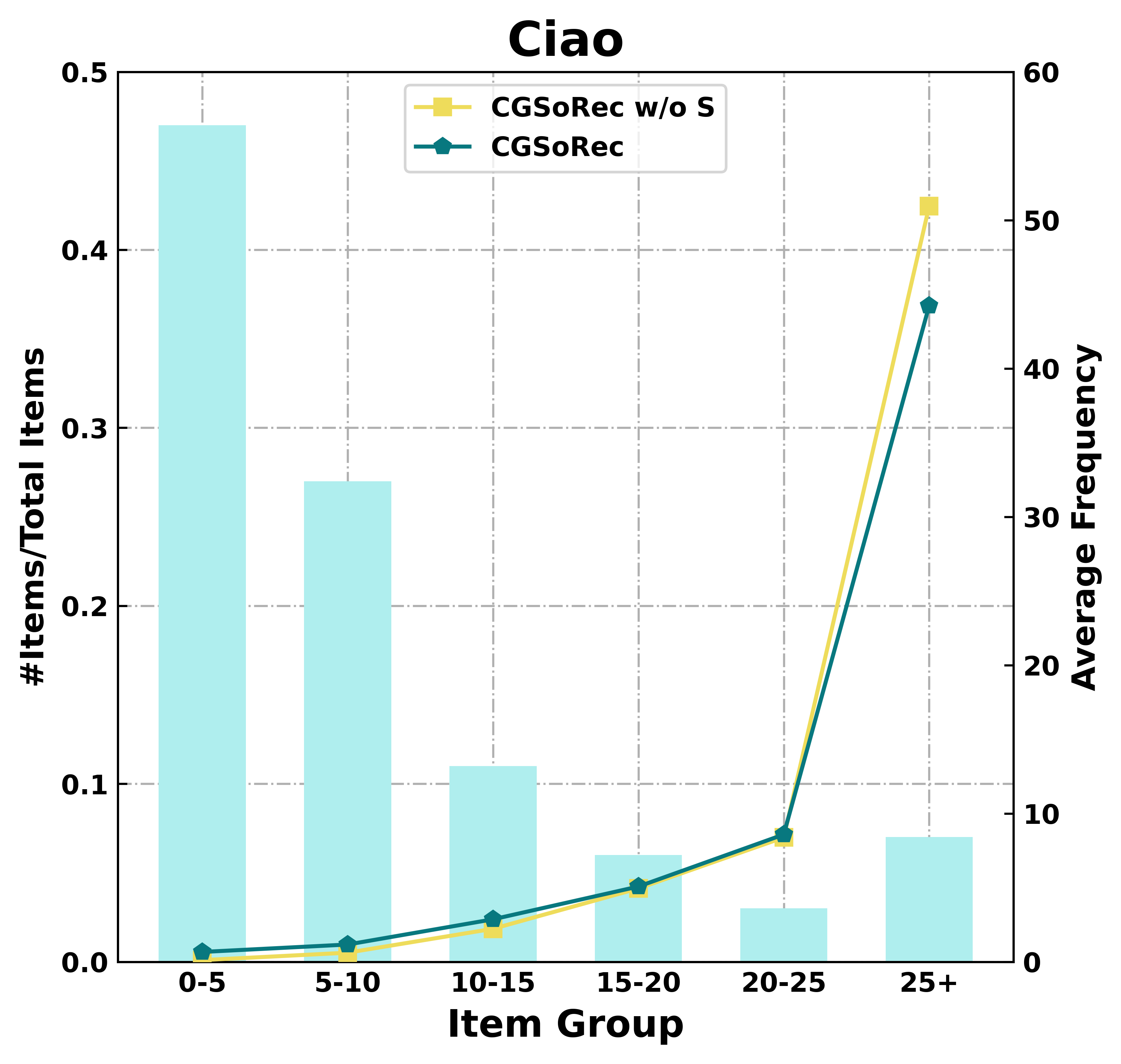}}}

\caption{Frequency of different item groups recommended by DiffNet, LightGCN-S, MHCN and CGSoRec before and after the incorporation of social networks in LastFM and Ciao dataset.}\label{fig:popular_bias_changes}
\end{figure*}

Table~\ref{tab:performance_com} and Fig.~\ref{fig:popular_bias_changes} reports the overall performance and popularity bias changes of compared models before and after the incorporation of social networks, separately.
According to the results, we can have the following observations: 
\begin{itemize}
    \item Existing recommendation models including social network like DiffNet,  MHCN, LightGCN-S and DiffRec show a decline in performance when social information is incorporated. This is primarily because the social network amplifies the popularity bias of the recommendation model. Consequently, the model tends to recommend hot items to users while reducing the frequency of long-tail items.
    \item Our proposed \textbf{CGSoRec} exhibits significant performance improvements across three datasets following the integration of social networks. For instance, on the LastFM dataset, Recall@5 increased from 0.0188 to 0.0227, an enhancement exceeding 20\%. 
    These improvements demonstrate the efficacy of our social denoising and the adjustment of user social preference weights.
    \item The performances of MHCN and GDMSR are lower than that of LightGCN across three datasets. 
    This could be attributed to the hypergraph convolution module in them, which intensifies the model's popularity bias.
    As a result, even after removing the self-supervised loss, the popularity bias of MHCN and GDMSR remains higher than that of LightGCN.
    \item The performance of the debiased social enhancement method MACR-S surpasses that of our proposed method, CGSoRec, on certain datasets. This is attributed to MACR's use of causal learning techniques to reduce popularity bias in recommendation models. However, when social networks are incorporated into the MACR model, its performance experiences a notable decline. In contrast, the performance of our CGSoRec model improves upon the integration of social networks, which effectively demonstrates the efficacy of our approach.
    \item All baseline methods (DiffNet, LightGCN-S, MHCN, MACR-S and DiffRec-S) exhibit an increased recommendation frequency for hot items following the introduction of the social network. 
    This suggests that the social network indeed amplifies the popularity bias in the recommendation model.
    In contrast, our proposed model \textbf{CGSoRec} demonstrates a decrease in the recommendation frequency of hot items following the introduction of social networks. This is attributed to our innovative approach of social denoising and the adjustment of users' social preference weights.
\end{itemize}

\subsection{Performance comparison in hot and long-tail items~(RQ2)}
In this subsection, we examine the performance improvements of the proposed model achieved for both hot and long-tail items. Based on the number of interact (Top 2\% - 7\% frequently interacted and rest) in the training set, the item groups are divided into the hot item group and the long-tail item group.

We present the recommendation performance of two distinct item groups on the LastFM, DBook and Ciao datasets under MACR-S, DiffRec-S and our proposed method CGSoRec in Table~\ref{tab:long_and_tail_performance_com}.
From the Table~\ref{tab:long_and_tail_performance_com}, we can see that with the introduction of social networks, the debiased recommendation mtehod MACR enhances recommendation performance for hot items while decreasing performance for long-tail items. 
The primary reason for this phenomenon is that the social network makes MACR tend to recommend more hot items and fewer long-tail items to users.
In addition, we observe that with the integration of social networks, the diffusion recommendation method DiffRec experiences a decline in recommendation performance for both hot and long-tail items. 
This is mainly due to the DiffRec method directly incorporates users' social preferences into the diffusion process as a condition, which is highly susceptible to noise within the social network. Consequently, even though the social network allows DiffRec to recommend more popular items, the noise in the network may impair the accuracy of these recommendations, thereby reducing the recommendation performance for hot items.
Compared to DiffRec, our proposed method CGSoRec, upon incorporating social networks, trades a minor decline in recommendation performance for hot items for a significant improvement in the recommendation performance of long-tail items.
As evident from the experimental results, CGSoRec traded a marginal reduction (3.67\% under the Recall@10 metric on dataset Lastfm and 1.17\% under the NDCG@10 metric on dataset DBook) in the recommendation performance of hot items for a substantial improvement (22.61\% under the Recall@10 metric on dataset Lastfm and 11.90\% under the NDCG@10 metric on dataset DBook) in the recommendation performance of long-tail items. 
This suggests that CGSoRec can effectively balance recommendation performance between hot and long-tail items and successfully mitigate the model's popular bias through social denoising and the adjustment of users’ social preference weights.

\subsection{Ablation Studies}
\subsubsection{Effect of Social Denoising and Weight Adjustment (RQ3)} To further study the effect of the Social Denoising and Weight Adjustment, we design two variants of our proposed \textbf{CGSoRec}: (1) \textbf{CGSoRec-SD}, where Condition-Guided Social Denoising Model (CSD) has been removed, and (2) \textbf{CGSoRec-SD\&WA}, where Social Denoising and Weight Adjustment both have been removed.
The experimental results are given in Table~\ref{tab:Ablation}. 
We can see that Removing Social Denoising leads to a slight decrease in the model's performance. 
For example, in the LastFM dataset, after removing Social Denoising, the model's NDCG@10 decreases about 1.6\%. 
This maybe due to Social Denoising removing redundant social relationships in social networks, allowing the model to more accurately reflect users' social preferences. 
In contrast, the removal of Weight Adjustment has a more significant impact on the model's performance. For example, in the Dbook dataset, after removing Weight Adjustment, the model's Recall@10 decreases about 23.73\%. 
This could be because Weight Adjustment mitigates the model's popularity bias by adjusting the weights of long-tail and popular items in users' social preferences. Removing Weight Adjustment deprives the model of its ability to debias the recommendation results, thus leading to a substantial decrease in the model's performance.

\begin{table*}[!tp]
\centering
\caption{Effect of Social Denoising and Weight Adjustment.}
\label{tab:Ablation}
\scalebox{0.65}{
\begin{tabular}{ccccccccccc}
\toprule
\multicolumn{2}{c}{\textbf{Datasets}}              & 
\multicolumn{3}{c}{\textbf{LastFM}} &\multicolumn{3}{c}{\textbf{Dbook}} & \multicolumn{3}{c}{\textbf{Ciao}}\\ 
\multicolumn{1}{c}{\textbf{Basemodel}}              & \multicolumn{1}{c}{\textbf{Method}} &
\multicolumn{1}{c}{\textbf{R@5}} & \multicolumn{1}{c}{\textbf{R@10}} &\multicolumn{1}{c}{\textbf{N@10}}&
\multicolumn{1}{c}{\textbf{R@5}} & \multicolumn{1}{c}{\textbf{R@10}} &\multicolumn{1}{c}{\textbf{N@10}}&
\multicolumn{1}{c}{\textbf{R@5}} & \multicolumn{1}{c}{\textbf{R@10}} &\multicolumn{1}{c}{\textbf{N@10}}\\
\midrule
\multirow{3}{*}{\textbf{CGSoRec}}
& \textbf{Ori.} &    \textbf{0.0227} &  \textbf{0.0491}    &   \textbf{0.0442}  &\textbf{0.0065}&\textbf{0.0121}&\textbf{0.0150}&\textbf{0.0031}&\textbf{0.0057}&\textbf{0.0057} \\  
& \textbf{w/o SD} & 0.0218  & 0.0476  & 0.0435&0.0062&0.0118&0.0144&0.0029&0.0054&0.0055   \\
& \textbf{w/o SD\&WA}  & 0.0171 & 0.0380 &  0.0365&0.0049&0.0090&0.0113&0.0025&0.0045& 0.0046\\ 
    \bottomrule
\end{tabular}}
\end{table*}

\begin{figure*}[!t]
\centering
{\subfigure[LastFM]
{\includegraphics[width=0.32\linewidth]{{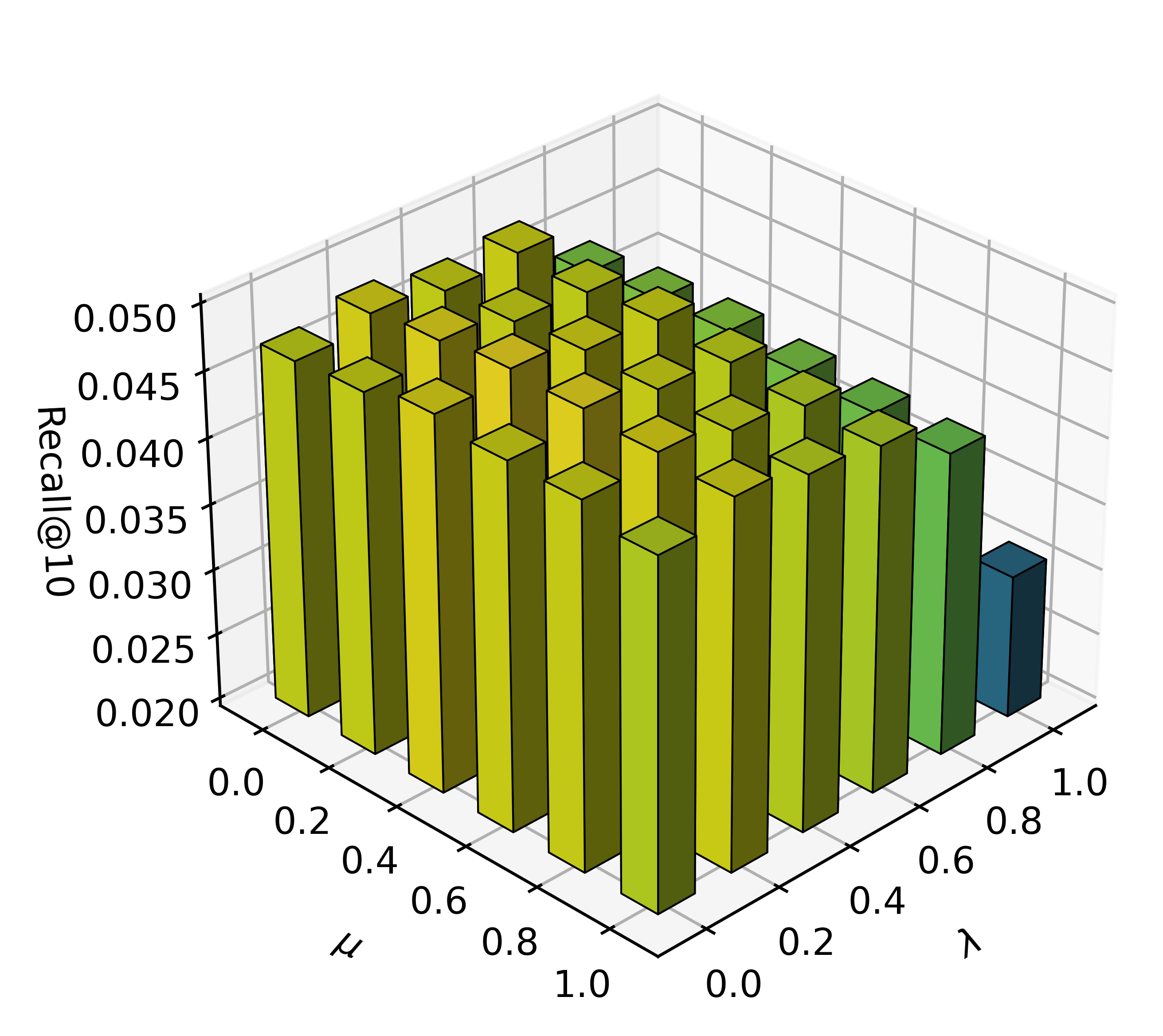}}}}
{\subfigure[Dbook]
{\includegraphics[width=0.32\linewidth]{{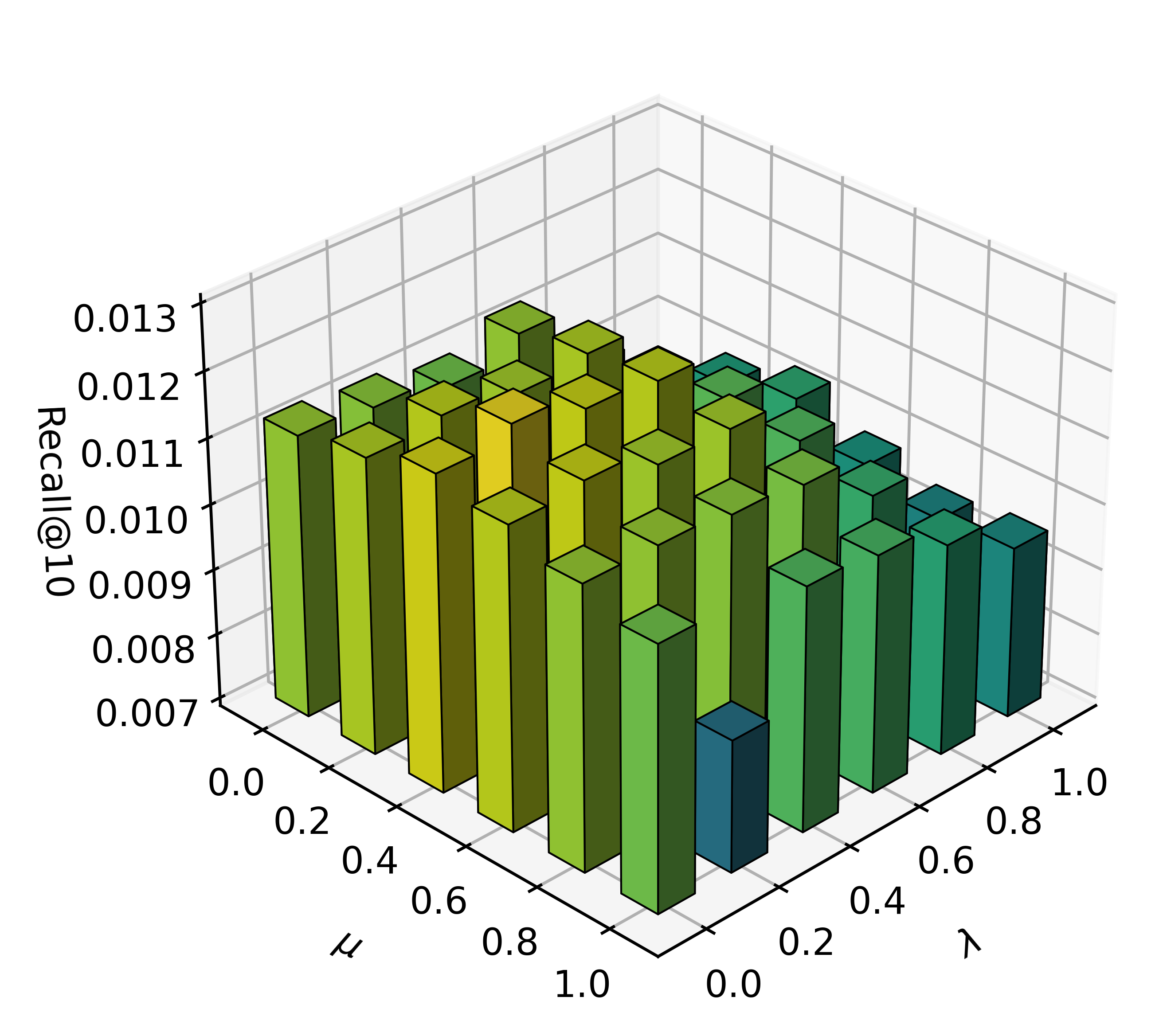}}}}
{\subfigure[Ciao]
{\includegraphics[width=0.32\linewidth]{{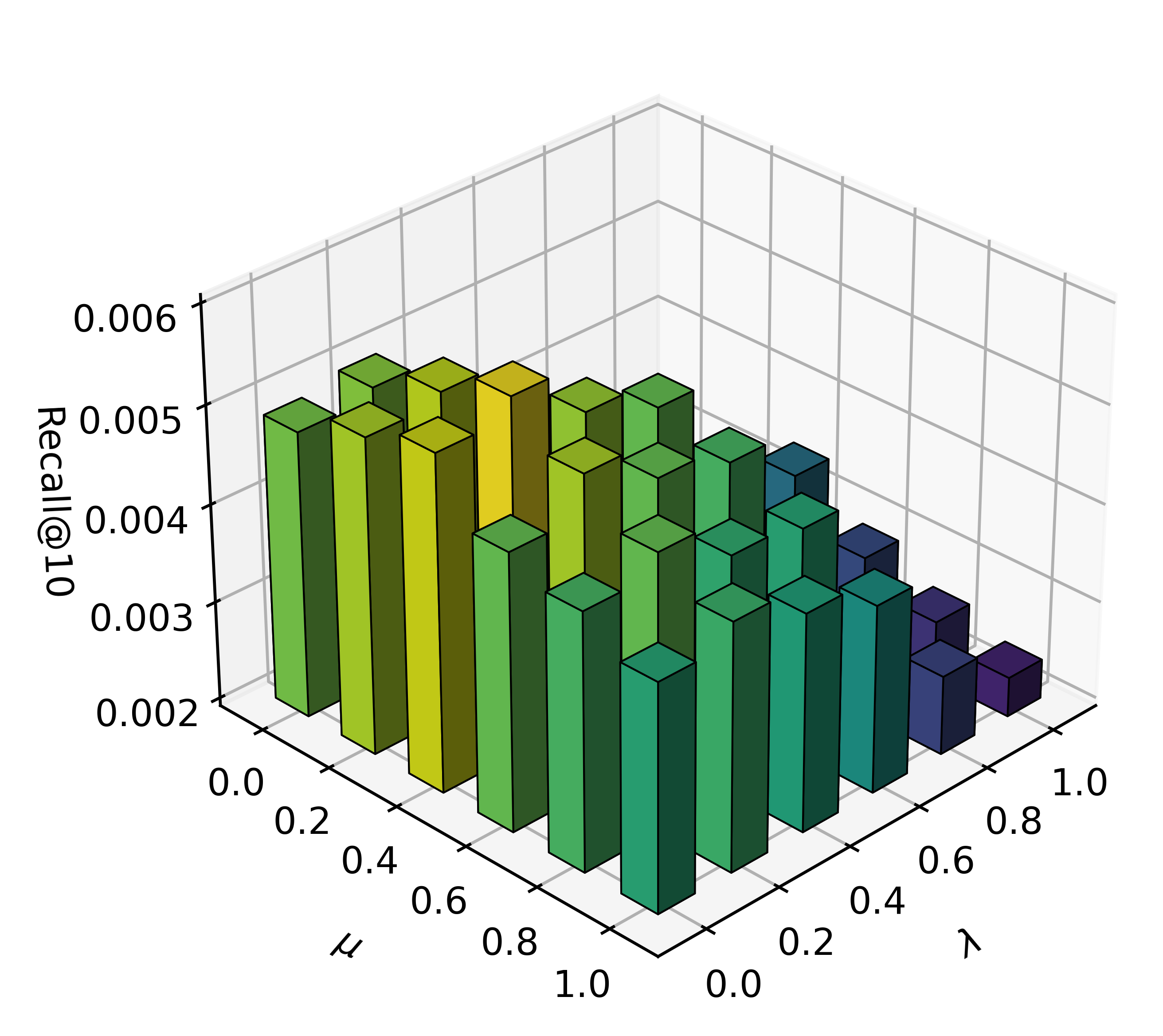}}}}%
\caption{Effect of $\mu$ and $\lambda$ on CGSoRec  w.r.t Recall@10.}

\label{fig:effect_of_hyperparameter_mu_lambda}
\end{figure*}

\subsubsection{Effect of hyper-parameter $\mu$ and $\lambda$ (RQ4)}

In this subsection, we analyze the impact of the hyper-parameters $\mu$ and $\lambda$ on the model performance. Figure~\ref{fig:effect_of_hyperparameter_mu_lambda} shows the variation in model performance with respect to $\mu$ and $\lambda$ across three datasets, from which we observe that the choice of the two hyper-parameters is crucial for the performance of CGSoRec. Smaller values of $\mu$ and $\lambda$ often lead to better model performance.
Overall, CGSoRec exhibits robustness to these two hyper-parameters on the LastFM dataset. Although the optimal choice of hyper-parameters varies across different datasets, larger values of $\mu$ and $\lambda$ generally lead to a sharp decline in performance. This phenomenon occurs across all three datasets. In our experiments, we selected the optimal combination of these two hyper-parameters to achieve the best experimental results.

\subsubsection{Effect of hyper-parameter $\omega_r$ (RQ4)} 
\begin{figure*}[h]
\centering
{\subfigure[LastFM]
{\includegraphics[width=0.32\linewidth]{{ 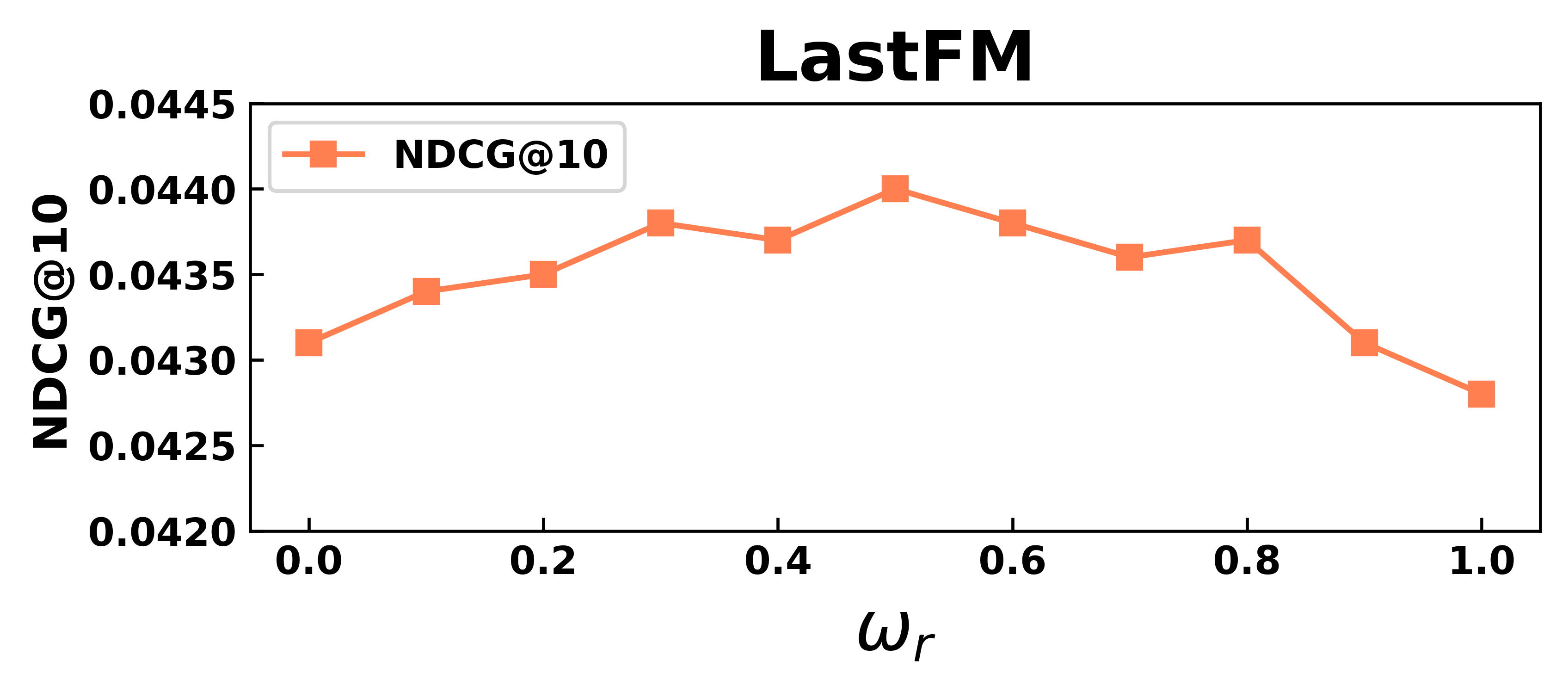}}}}
{\subfigure[Dbook]
{\includegraphics[width=0.32\linewidth]{{ 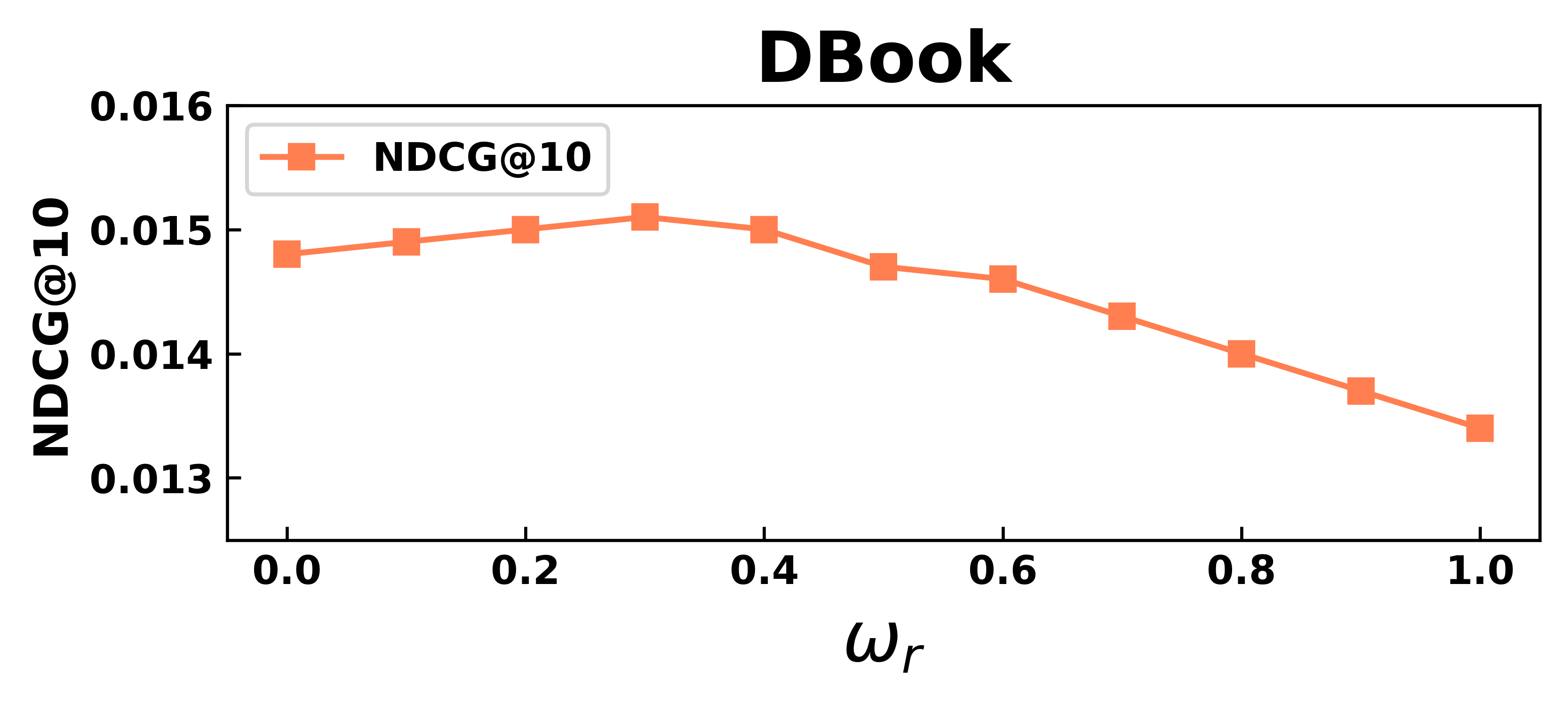}}}}
{\subfigure[Ciao]
{\includegraphics[width=0.32\linewidth]{{ 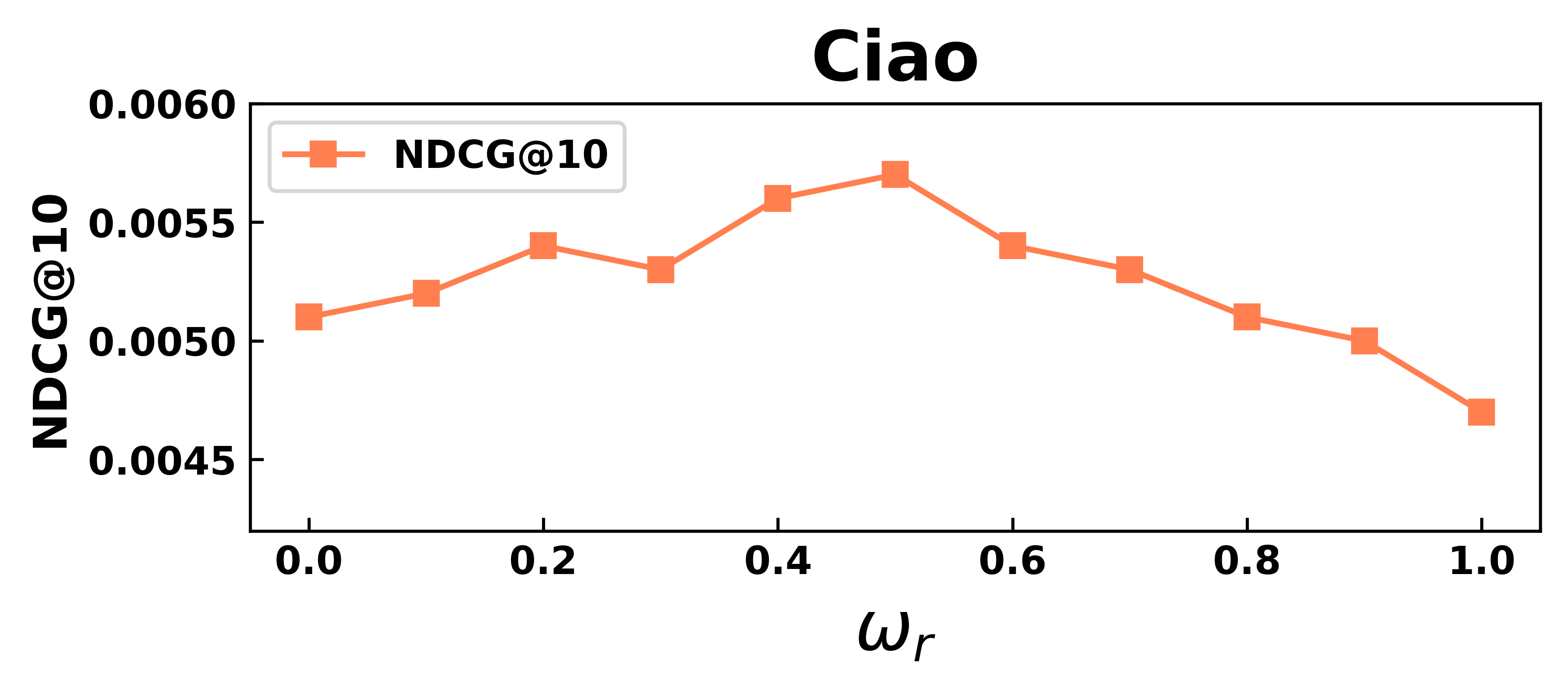}}}}%
\caption{Effect of $\omega_r$ on CGSoRec  w.r.t NDCG@10.}

\label{fig:effect_of_hyperparameter}
\end{figure*}

The hyper-parameter $\omega_r$ as formulated in Eq.~\ref{final_x} controls the degree to which the condition-guided recommendation result is involved in the final prediction.
We execute a series of experiments on the LastFM, DBook and Ciao dataset using our proposed \textbf{CGSoRec} model, assessing its performance specifically in terms of NDCG@10. 
The experimental results are given in Fig.~\ref{fig:effect_of_hyperparameter}. 
We can see that, in lastFM and Ciao dataset, as the value of $\omega_r$ varies from 0 to 0.5, the model's performance progressively improves. However, further increasing $\omega_r$ beyond this point is counterproductive.
Similarly, in DBook dataset, as the value of $\omega_r$ varies from 0 to 0.3, the model's performance progressively improves. And further increasing $\omega_r$ beyond this point is counterproductive.
This observation indicates that the proper integration of conditional model recommendation results is advantageous in mitigating model popularity bias and enhancing overall model performance.

\begin{figure*}[h]
\centering
{\subfigure[LastFM]
{\includegraphics[width=0.32\linewidth]{{ 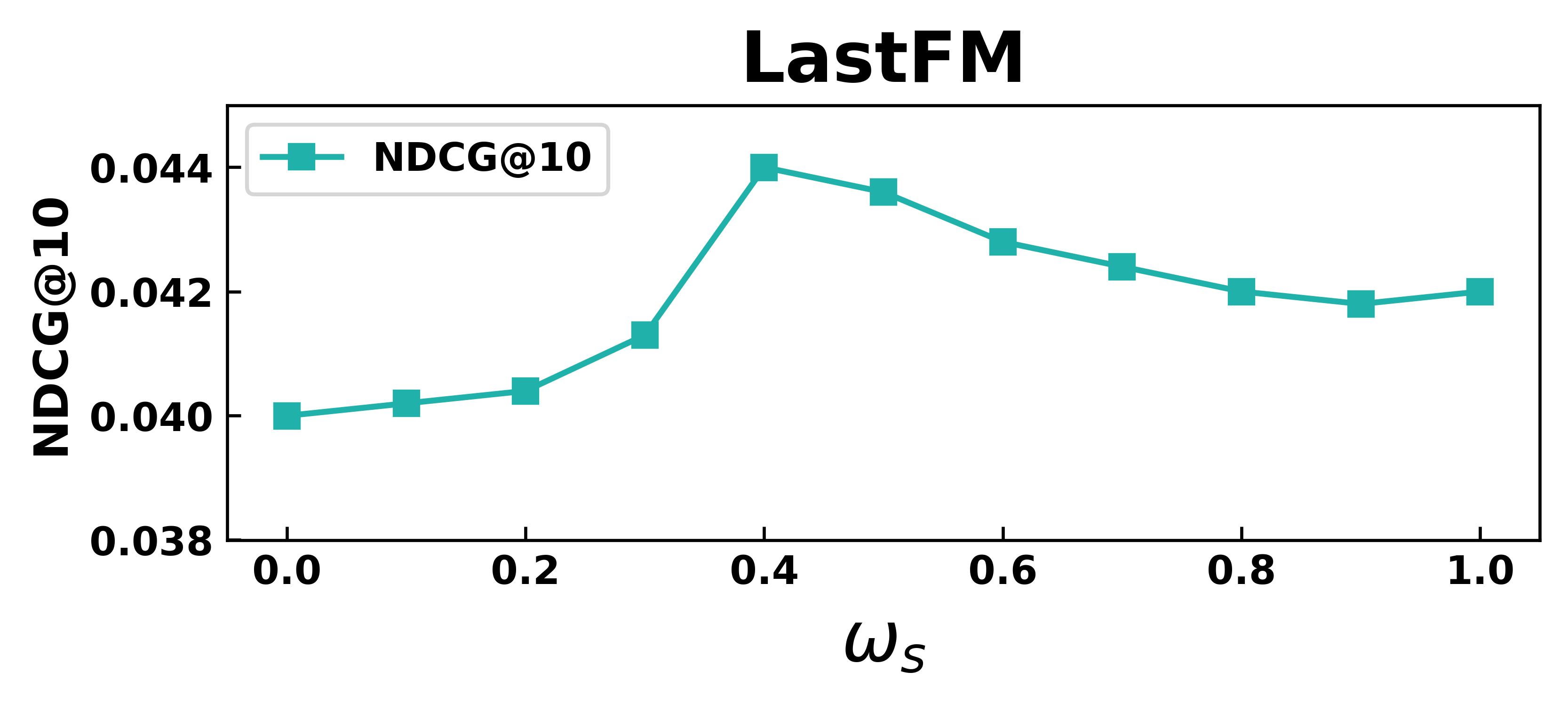}}}}
{\subfigure[Dbook]
{\includegraphics[width=0.32\linewidth]{{ 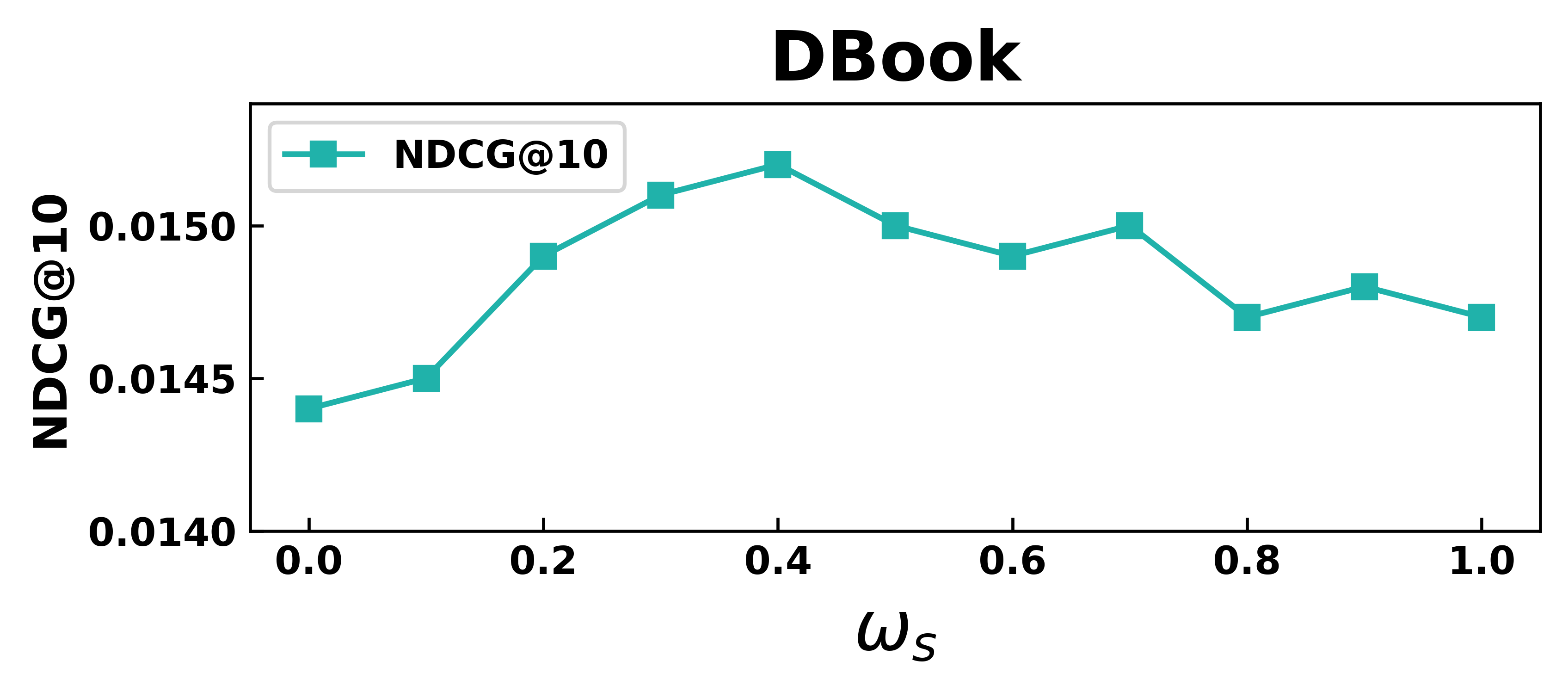}}}}
{\subfigure[Ciao]
{\includegraphics[width=0.32\linewidth]{{ 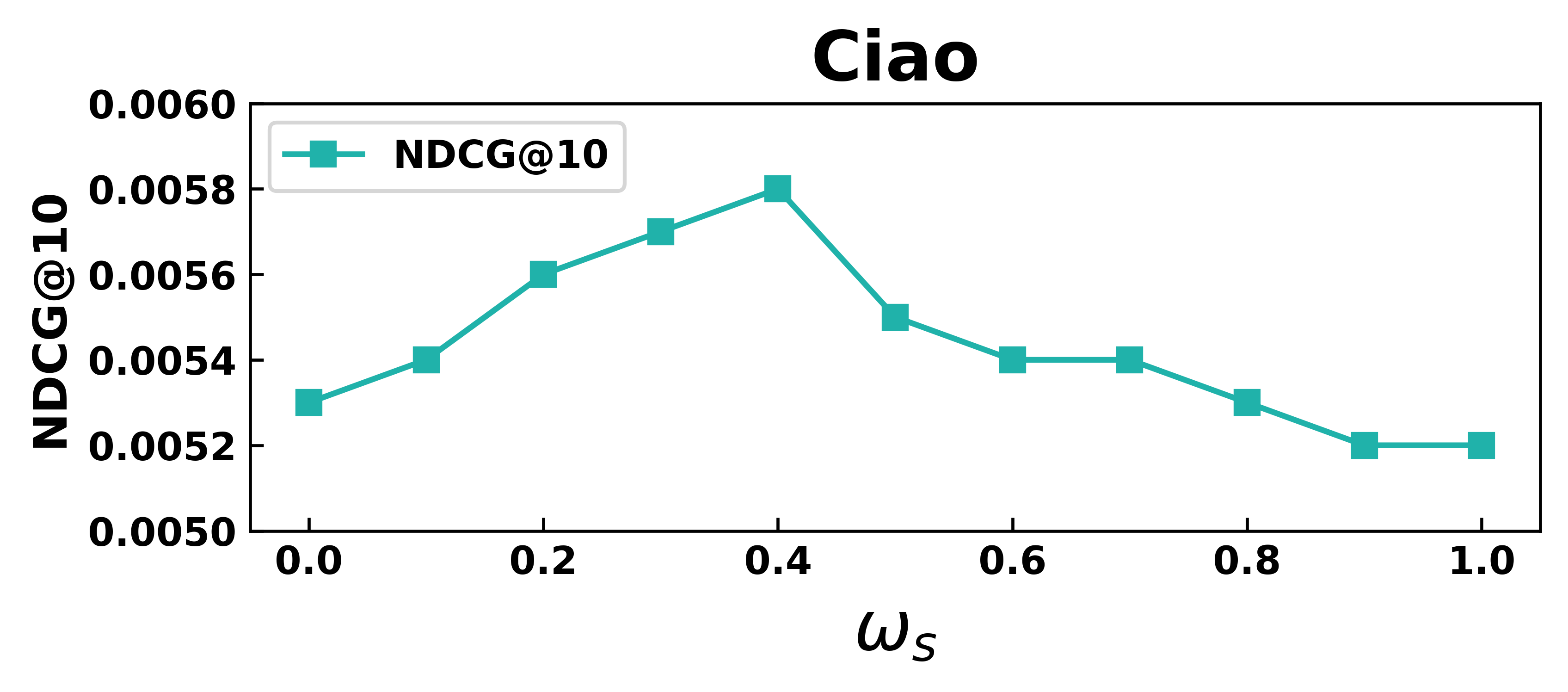}}}}%
\caption{Effect of $\omega_s$ on CGSoRec  w.r.t NDCG@10.}

\label{fig:effect_of_hyperparameter_sw}
\end{figure*}

\subsubsection{Effect of hyper-parameter $\omega_s$ (RQ4)} The hyper-parameter $\omega_s$ as formulated in Eq.~\ref{final_s} controls the degree to which the condition-guided recommendation result is involved in the final denoised result of social network. 
A larger value of parameter $\omega_s$ indicates that the social relationships between two users who have similar preferences for long-tail items are more likely to be preserved during the denoising process. 
Additionally, as $\omega_s$ increases, the conditions will also dominate the final denoising result, leading to a decrease in the denoising capability of the model.
We also execute a series of experiments on the LastFM, DBook and Ciao dataset using our proposed \textbf{CGSoRec} model, assessing its performance specifically in terms of NDCG@10. 
The experimental results are given in Fig.~\ref{fig:effect_of_hyperparameter_sw}. 
In LastFM, DBook and Ciao dataset, as the value of $\omega_s$ varies from 0 to 0.4, the model's performance progressively improves. And when we further increase $\omega_s$ beyond this point is counterproductive.
This phenomenon suggests that appropriate conditions are also necessary to guide the denoising process of social networks, in order to obtain denoising results that are conducive to mitigate popularity bias in recommendation model.

\section{Conclusion}
In this work, we proposed a Condition-Guided Social Recommendation Model (\textbf{CGSoRec}), which can mitigate the popularity bias within the recommendation model by the adjusted social performance.
Our model offers advanced components that yield two key advantages: 
(1) The Condition-Guided Social Denoising Model (CSD) included in \textbf{CGSoRec} can remove redundant or noisy social relations in social networks for more accurate user social preferences.
(2) The Condition-Guided Diffusion Recommendation Model (CGD) included in \textbf{CGSoRec} can introduce the adjusted social performance (increasing the weight of long-tail items while decreasing the weight of hot items) as a condition to control the recommendation results. 
Comprehensive experiments conducted on three real-world datasets effectively demonstrate the efficacy of our proposed method.
The ablation studies also demonstrate that the two components of our model can mitigate the model's popularity bias and improve the recommendation performance.

\bibliographystyle{elsarticle-num}
\bibliography{sample-base}

\end{document}